\def\PRL{{ Phys. Rev. Lett.\ }\/}
\def\PRB{{ Phys. Rev. B\ }\/}
\def\PRX{{ Phys. Rev. X\ }\/}

\def\etal{{\it et al.~}\/}

\def\be{\begin {equation}}
\def\ee{\end {equation}}
\def\ber{\begin {eqnarray}}
\def\eer{\end {eqnarray}}
\def\bers{\begin {eqnarray*}}
\def\eers{\end {eqnarray*}}

\makeatletter

\newcommand{\Rmnum}[1]{\expandafter\@slowromancap\romannumeral #1@}

\makeatother

\makeatletter
\newcommand*\env@matrix[1][*\c@MaxMatrixCols c]{%
  \hskip -\arraycolsep
  \let\@ifnextchar\new@ifnextchar
  \array{#1}}
\makeatother

\documentclass[aps,prb,showpacs,superscriptaddress,a4paper,twocolumn]{revtex4-1}
\usepackage{amsmath}
\usepackage{float}
\usepackage{color}

\usepackage[normalem]{ulem}
\usepackage{soul}
\usepackage{amssymb}

\usepackage{amsfonts}
\usepackage{amssymb}
\usepackage{graphicx}
\begin {document}


\title{Unremovable linked nodal structures protected by crystalline symmetries in stacked bilayer graphene with Kekul\'{e} texture} 

\author{Chiranjit Mondal}

\author{Sunje Kim}

\author{Bohm-Jung Yang}
\email{bjyang@snu.ac.kr}

\affiliation{Center for Correlated Electron Systems, Institute for Basic Science (IBS), Seoul 08826, Korea}
\affiliation{Department of Physics and Astronomy, Seoul National University, Seoul 08826, Korea}
\affiliation{Center for Theoretical Physics (CTP), Seoul National University, Seoul 08826, Korea}

\date{\today}

\begin{abstract}
Linking structure is a new concept characterizing topological semimetals, 
which indicates the interweaving of gap-closing nodes at the Fermi energy ($E_F$) with other nodes below $E_F$.
As the number of linked nodes can be changed only via pair-creation or pair-annihilation, a linked node is more stable and robust than ordinary nodes without linking.
Here we propose a new type of a linked nodal structure between a nodal line (nodal surface) at $E_F$ with another nodal line (nodal surface) below $E_F$ in two-dimensional (three-dimensional) spinless fermion systems with $\mathcal{IT}$ symmetry where $\mathcal{I}$ and $\mathcal{T}$ indicate inversion and time-reversal symmetries, respectively.
Because of additional chiral and rotational symmetries, in our system, a double band inversion creates a pair of linked nodes carrying the same topological charges, thus the pair are unremovable via a Lifshiftz transition, which is clearly distinct from the cases of the linked nodes reported previously.
A realistic tight binding model and effective theory are developed for such a linking structure. 
Also, using density functional theory calculations, 
we propose a class of materials, composed of stacked bilayer graphene with Kekul\'{e} texture, 
as a candidate system hosting the new type of the linked nodal structure. 
\end{abstract}


\maketitle

\par {\it  Introduction:} 
Topological semimetals and nodal superconductors indicate the gapless topological phases 
with nodal points (NPs)/ nodal loops (NLs)/ nodal surfaces (NSs) near the Fermi energy ($E_F$) \cite{W1,W2,W3,W4,W5,W6,W7,W8, W9,W10, D1,D2,D3,D4,D5,D6,D7,D8, NL1,NL2,NL3,NL5,NL6,NL7,NL8, T2,T3, TS1,TS2,TS3,TS4,TS5}.
Normally, such a node is characterized by its primary topological charge defined in a lowest-dimensional manifold enclosing
the node such as a two-dimensional (2D) surface/one-dimensional (1D) loop/zero-dimensional (0D) point enclosing a NP/NL/NS, respectively, in a three-dimensional (3D) momentum space \cite{RMP1, RMP2, AnnRev,NSS1,NSS2,NSS3,NSS4}.
Although the presence of a primary topological charge indicates the local stability of the relevant nodal structure,
it does not guarantee the global stability of the node \cite{DCN1, DCN2}. 
For example, a single NL or NS in 3D systems with a primary topological charge 
can be annihilated via a continuous deformation \cite{DCN1,DCN2,DCN3}.

Recently, it is found that there are a class of doubly charged (DC) nodes with two distinct topological charges, 
which are more robust than ordinary singly charged nodes~\cite{DCN1,DCN2,DCN3,DCN4,DCN5,DCN6,DCN7}.
Two important characteristics of DC nodes are as follows.
First, the number of DC nodes can be changed only via pair-creation or pair-annihilation,
that is, an annihilation of a single DC node is not allowed \cite{DCN1,DCN2}.
Second, a pair of DC nodes at $E_F$ are linked with another node below $E_F$ \cite{L1,L2,L3,L4,DCN7}.
For example, in 3D centrosymmetric systems belonging to the Altland-Zirnbauer (AZ) class AI and CI,
a doubly charge NL at $E_F$ is always linked with another NL below $E_F$ \cite{AZ,SF,L1,DCN7}.

More recently, it is found that a doubly charged NS at $E_F$ is linked with a NL below $E_F$
in AZ class BDI systems with inversion symmetry \cite{L2}.
The full list of such linking structures and their relation with DC nodes
were recently established\cite{L2} within the framework of the AZ classification including an additional inversion $\mathcal{I}$ symmetry,
dubbed the AZ+$\mathcal{I}$ classification~\cite{DCN3}. 
However, understanding the influence of additional crystalline symmetries, beyond the AZ+$\mathcal{I}$ classification scheme, on the DC nodes with linking structures is still an important open problem.

In this letter, we propose a new class of linked nodal structures between two concentric NLs (NSs) at $E_F$ and
another NL (NS) below $E_F$ in 2D (3D) spinless fermion systems belonging to the AZ+$\mathcal{I}$ class BDI with two extra symmetries: 
one is a three-fold rotation $C_{3z}$ about the $z$-axis, and the other is a chiral symmetry $\mathcal{C}^{\prime}$
that is different from the intrinsic chiral symmetry $\mathcal{C}$ of the class BDI.
As both $\mathcal{C}$ and $\mathcal{C}^{\prime}$ anticommute with the Hamiltonian,
their combination gives a commuting symmetry $\mathcal{O}_L \equiv i \mathcal{C} \mathcal{C}^{\prime}$ which is local in the momentum $\mathbf{k}$.
In 2D systems, $\mathcal{O}_L$ is nothing but the mirror $M_z:(x,y,z)\rightarrow(x,y,-z)$ symmetry about the basal plane of the system or a layer exchange symmetry of a 2D bilayer structure. 
When such 2D bilayers are vertically stacked with weak inter-bilayer coupling, $\mathcal{O}_L$ still remains as an excellent $\mathbf{k}$-local symmetry of the resulting 3D structure, which supports linked NSs. 
Interestingly, in this class of systems, a double band inversion (DBI) creates two concentric NLs or NSs at $E_F$ that carry two distinct 0D topological charges: one is $Z_2$ type ($Z^{0D}_2$) while the other is $Z$-type ($Z^{0D}$).
As the two NLs (NSs) at $E_F$ have the same $Z$-type charges, their pair-annihilation through a Lifshitz transition is forbidden.
Such a stability against the pair-annihilation of DC nodes does not exist in any other DC nodes known up to now.

We construct a realistic model and symmetry-based effective theory considering a Kekul\'{e} textured bi-layer graphene (BLG) and the related layered 3D structure as an example. 
Also, using the \emph{ab-initio} band structure calculations, we propose a class of materials with the chemical formula AC$_{12}$ (A =Zn,Al,Be) in which the interplay of two distinct chiral symmetries and $C_{3z}$ protects the linked cylindrical NSs near $E_F$.
We also show that breaking the chiral symmetries transforms the NSs to NLs as realized in materials BC$_{12}$ (B =Li,B,Mg) in which NLs are protected by $\mathcal{IT}$ and $M_z$ symmetry where $\mathcal{T}$ indicates time-reversal ($\mathcal{T}$) symmetry.

\begin{figure}[t!]
	\centering
	\includegraphics[width=0.5\textwidth]{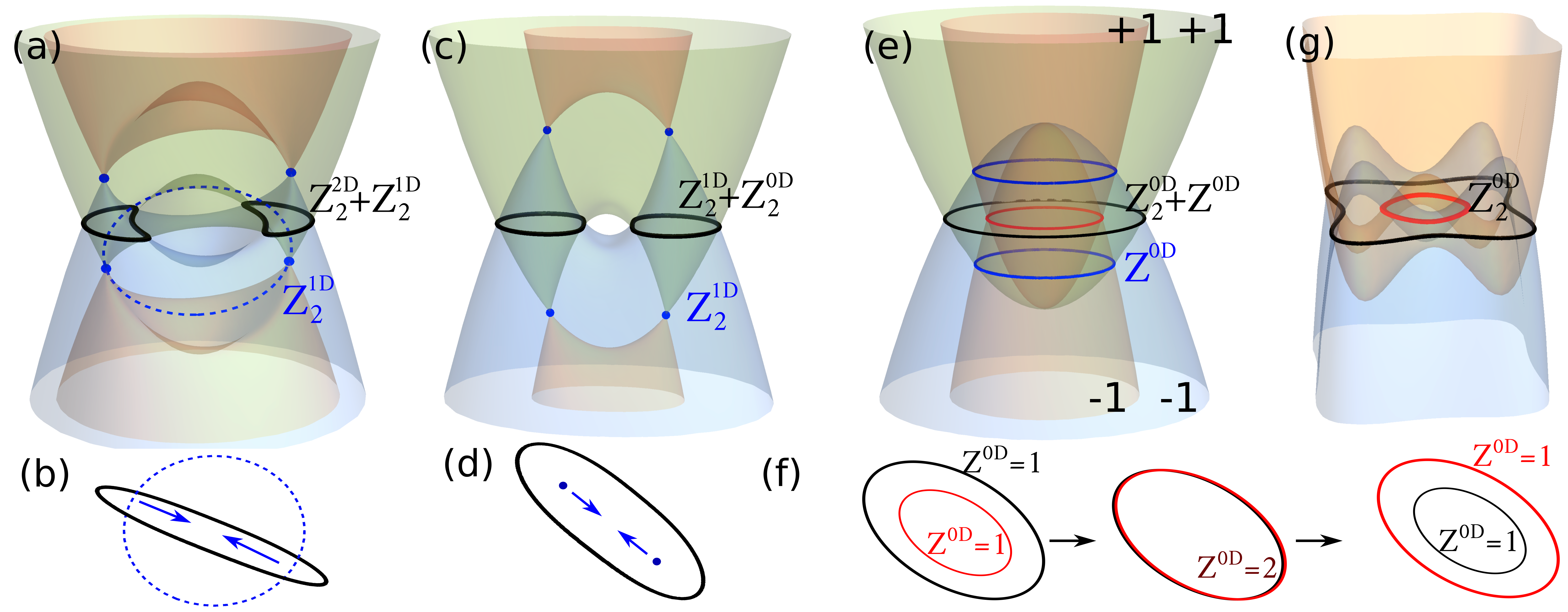}
	\caption{
		Doubly charged (DC) nodes and their linking structures.
		(a) Linked nodes in 3D systems belonging to the AZ+$\mathcal{I}$ class AI or CI. 
		A monopole NL at $E_F$ (black loop) is linked with another NL below $E_F$ (blue dotted loop).
		$Z^{dD}$ ($Z^{dD}_2$) ($d=0,1,2$) near each node indicates its $d$-dimensional $Z$ ($Z_2$) charge.
		The relevant nodal structure after merging of two monopole NLs is shown in (b).
		Both the merged NL at $E_F$ and the NL below $E_F$ can be annihilated.
		(c) Linked nodes in 2D systems belonging to the AZ+$\mathcal{I}$ class BDI.
		A NL at $E_F$ (black loop) always accompanies a NP below $E_F$ (blue dot).
		The relevant nodal structure after merging of two NLs is shown in (d).
		Both the merged NL at $E_F$ and two NPs below $E_F$ can be annihilated.
		(e) Linked nodes in 2D systems belonging to the AZ+$\mathcal{I}$ class BDI with an additional chiral and $C_{3z}$ symmetries.
		$\pm 1$ denote the $\mathcal{O}_L$ eigenvalues.
		Two concentric NLs at $E_F$ accompany another NL below $E_F$.  
		The merging process of two concentric NLs at $E_F$ is described in (f). 
		As they have the same $Z^{0D}$, pair-annihilation is impossible.
		(g) Two concentric NLs generated by a single band inversion. 
		Two loops can be pair-annihilated.
	}
	\label{DBI}
\end{figure}

{\it Double band inversion (DBI) and linking structure.|}  
A pair of DC nodes with linking structure can be created via a DBI process
in which two valence and two conduction bands are simultaneously inverted.
For instance, in a 3D system belonging to the AZ+$\mathcal{I}$ class AI or CI, a DBI creates a pair of monopole NLs at $E_F$
that carry both a 1D $Z_2$ charge $Z^{1D}_2$ (equivalent to the $\pi$ Berry phase) and a 2D $Z_2$ monopole charge $Z^{2D}_2$
as described in Fig.~\ref{DBI}(a)~\cite{L1}.
Each monopole NL at $E_F$ (black loop) is always linked with another NL below $E_F$ (blue dashed loop)~\cite{L1}.
When the band structure is smoothly deformed in a way that two monopole NLs merge and turn into a single trivial NL without monopole charge,
both the trivial NL at $E_F$ and the NL below $E_F$ can be continuously shrunk to a point, and then be annihilated [see Fig.~\ref{DBI}(b)].
A similar deformation is also possible in 2D systems belonging to the AZ+$\mathcal{I}$ class BDI 
where a NL at $E_F$ is linked with a NP below $E_F$~\cite{L2} [see Fig.~\ref{DBI}(c)].
When two DC NLs at $E_F$ are merged, both the merged NL at $E_F$ and the NP pair below $E_F$ can be annihilated
as described in Fig.~\ref{DBI}(d).

On the other hand, when an additional chiral symmetry and $C_{3z}$ are supplemented to the 2D class BDI systems,
a distinct type of DC NLs with extra stability can be created via a DBI.
In this case, each of the NLs at $E_F$ carries two distinct 0D charges ($Z^{0D}_2$ and  $Z^{0D}$),
and the two NLs at $E_F$ are linked with another NL below $E_F$ as shown in Fig.~\ref{DBI}(e). 
Interestingly, when the two concentric NLs at $E_F$ are merged, as they have the same $Z^{0D}$ charge,
the merged NL also carries a nonzero topological charge so that it cannot be annihilated. 
Thus further deformation splits the merged NL into two DC concentric NLs again as described in Fig.~\ref{DBI}(f).
We note that the emergence of concentric DC NLs is a direct consequence of a DBI.
Namely, when two concentric NLs are created by a single band inversion between one valence and one conduction bands as in Fig.~\ref{DBI}(g),
the merging of two concentric NLs always leads to their pair-annihilation and gap-opening.

\begin{figure}[t!]
	\centering
	\includegraphics[width=0.5\textwidth]{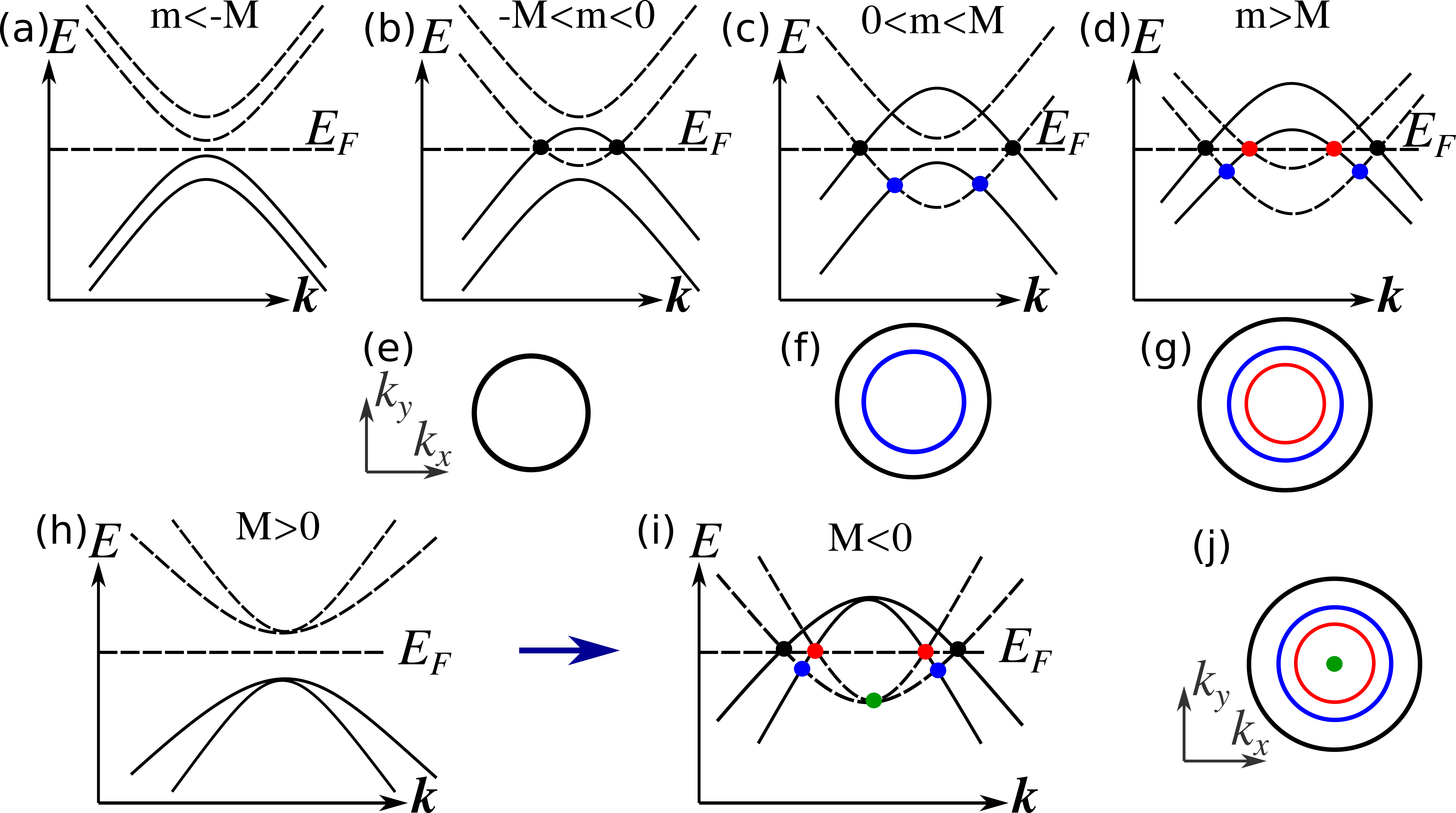}
	\caption{
		(a,b,c,d) A double band inversion (DBI) process in systems belonging to the AZ+$\mathcal{I}$ class BDI with an additional chiral symmetry $\mathcal{C}^{\prime}$.
		(e,f,g) The nodal structure corresponding to each upper panel.
		(h,i) BDI process when extra $C_{3z}$ symmetry is included.
		(j) The nodal structure corresponding to (i).
		The dashed (solid) lines indicate the bands carrying $+1$ ($-1$) $\mathcal{O}_L$ eigenvalues.
	}
	\label{model}
\end{figure}

{\it Continuum Hamiltonian.|}  
A general DBI process of 2D systems belonging to the AZ+$\mathcal{I}$ class BDI can be described by the following 4-band continuum Hamiltonian.
\begin{equation}\label{kp}
	H(\mathbf{k})= p(\mathbf{k}) \sigma_z \tau_0 + q(\mathbf{k}) \sigma_z \tau_x + r(\mathbf{k}) \sigma_z \tau_z + s(\mathbf{k}) \sigma_y \tau_y, 
\end{equation}
where the functions $p(\mathbf{k})$, $q(\mathbf{k})$, $r(\mathbf{k})$, $s(\mathbf{k})$ are even in the momentum $\mathbf{k}$.
Using the Pauli matrices $\tau_{x,y,z}$, $\sigma_{x,y,z}$, $2\times2$ identity matrices $\sigma_0$, $\tau_0$, and the complex conjugation operator $K$, we choose the symmetry representation $\mathcal{I}=\sigma_0 \tau_0$, $\mathcal{T}=K \sigma_0 \tau_0$, and $\mathcal{C}=\sigma_x \tau_0$,
which satisfy $(\mathcal{IT})^2=\mathcal{C}^2=1$. 
Also the combination of $\mathcal{IT}$ and $\mathcal{C}$ gives a $\mathbf{k}$-local particle-hole symmetry $\mathcal{P}=\mathcal{ITC}$.
Then the Hamiltonian satisfies, 
$(\mathcal{IT})H(\mathbf{k})(\mathcal{IT})^{-1}=H(\mathbf{k})$, $\mathcal{P}H(\mathbf{k})\mathcal{P}^{-1}=-H(\mathbf{k})$, and $\mathcal{C}H(\mathbf{k})\mathcal{C}^{-1}=-H(\mathbf{k})$,
which are the defining property of the AZ+$\mathcal{I}$ class BDI. 
We also impose $\mathcal{I}$ symmetry as the DBI occurs at the $\Gamma$ point.
When an additional chiral symmetry $\mathcal{C}^{\prime}=\sigma_y \tau_0$ exists, we find $s(\mathbf{k})=0$ and  $[H(\mathbf{k}),\mathcal{O}_L]=0$ where $\mathcal{O}_L = i \mathcal{C} \mathcal{C}^{\prime}$.

To describe a DBI process, we assume $p(\mathbf{k})=M$, $q(\mathbf{k})=k^2-m$, $r(\mathbf{k})=s(\mathbf{k)}=0$
where $M>0$ and $m$ are constants. 
As shown in Fig.~\ref{model}, when $m<-M$, we have a gapped band structure. 
When $-M<m<0$, a NL appears at $E_F$. 
When $0<m<M$, another NL emerges below $E_F$. 
Finally, when $m>M$, another NL at $E_F$ is generated, thus we have two concentric NLs at $E_F$ linked with another NL below $E_F$. 
We see that although a linked nodal structure can be created via a sequence of BDI process described in Fig.~\ref{model}(a,b,c,d),
the NLs at $E_F$ are created individually, not pairwise.
However, when $C_{3z}$ symmetry that imposes double degeneracy at $\mathbf{k}=0$ is supplemented,
the DBI process for the pair-creation of concentric NLs can be completed as shown in Fig.~\ref{model}(h,i,j).

To implement $C_{3z}$ symmetry, we take a matrix representation
$C_{3z}=e^{i \frac{2\pi}{3} \sigma_0 \tau_y}$ satisfying $(C_3)^3=1,~[C_3,T]=0$. 
The $C_{3z}$ invariance of the Hamiltonian can be satisfied by choosing
$p(\mathbf{k})=M+2\alpha (k_{x}^2+k_{y}^{2})$, $q(\mathbf{k})=-2\alpha k_x k_y$, $r(\mathbf{k})=\alpha (k_{x}^{2}-k_{y}^{2})$.
Assuming $\alpha>0$, we have a gapped band structure when $M>0$ as shown in Fig.~\ref{model}(h)
in which double degeneracy occurs at $\Gamma$ because the bands have complex $C_{3z}$ eigenvalues $\lambda=e^{i2\pi/3}$ and $\lambda^{*}$.
One can also show that the $\mathcal{O}_L$ eigenvalues of the unoccupied (occupied) bands are +1 (-1),
and all four bands have the same inversion eigenvalues at $\Gamma$.
When $M<0$, two NLs appear at $E_F$ with radii $\sqrt{|M|/\alpha}$ and $\sqrt{|M|/3\alpha}$, respectively. 
Between them, another NL with a radius $\sqrt{|M|/2\alpha}$ appears below $E_F$.

{\it Topological charges.|}  
Let us define $Z^{0D}$ and $Z^{0D}_2$, explicitly.
First, $Z^{0D}$ is defined by the eigenvalues ($\pm 1$) of $\mathcal{O}_L$ symmetry.
At a given momentum $\mathbf{k}$, we denote the number of occupied bands with the positive (negative) $\mathcal{O}_L$ eigenvalues by $\eta_+(\mathbf{k})$ ($\eta_-(\mathbf{k})$). Also we define $\Delta\eta(\mathbf{k}) = \eta_+(\mathbf{k}) - \eta_-(\mathbf{k})$.
For a given NL, we pick two momenta $\mathbf{k}_{\text{in}}$ and $\mathbf{k}_{\text{out}}$, which are inside and outside
of the loop, respectively. Then we define $Z^{0D} \equiv \frac{1}{2} \left[\Delta\eta(\mathbf{k}_{\text{out}})-\Delta\eta(\mathbf{k}_{\text{in}})\right]$.
One can easily show that $Z^{0D}=+1$ for both NLs at $E_F$ in Fig.~\ref{model}(d, i).
As they have the same topological charge, their pair-annihilation via merging is impossible.
 
Another charge $Z^{0D}_2$ is defined as follows~\cite{DCN3}.
Due to chiral symmetry, $H(\mathbf{k})$ can take a block off-diagonal form as $H(\mathbf{k})=\begin{pmatrix} 0&A(\mathbf{k})\\A^{T}(\mathbf{k})&0 \end{pmatrix}$ where $A(\mathbf{k})$ denotes a real matrix.
Then $Z^{0D}_2$ is defined as
\begin{align}
	Z^{0D}_2=\mathrm{sign}\{\mathrm{det}A(\mathbf{k}_{\text{in}})\cdot \mathrm{det}A(\mathbf{k}_{\text{out}})\}.
	\label{0D_BDI}
\end{align}
It is straightforward to show that the two concentric NLs at $E_F$ in Fig.~\ref{model}(d, i)
have the opposite $Z^{0D}_2$ charges.

\begin{figure}[t!]
	\centering
	\includegraphics[width=0.5\textwidth]{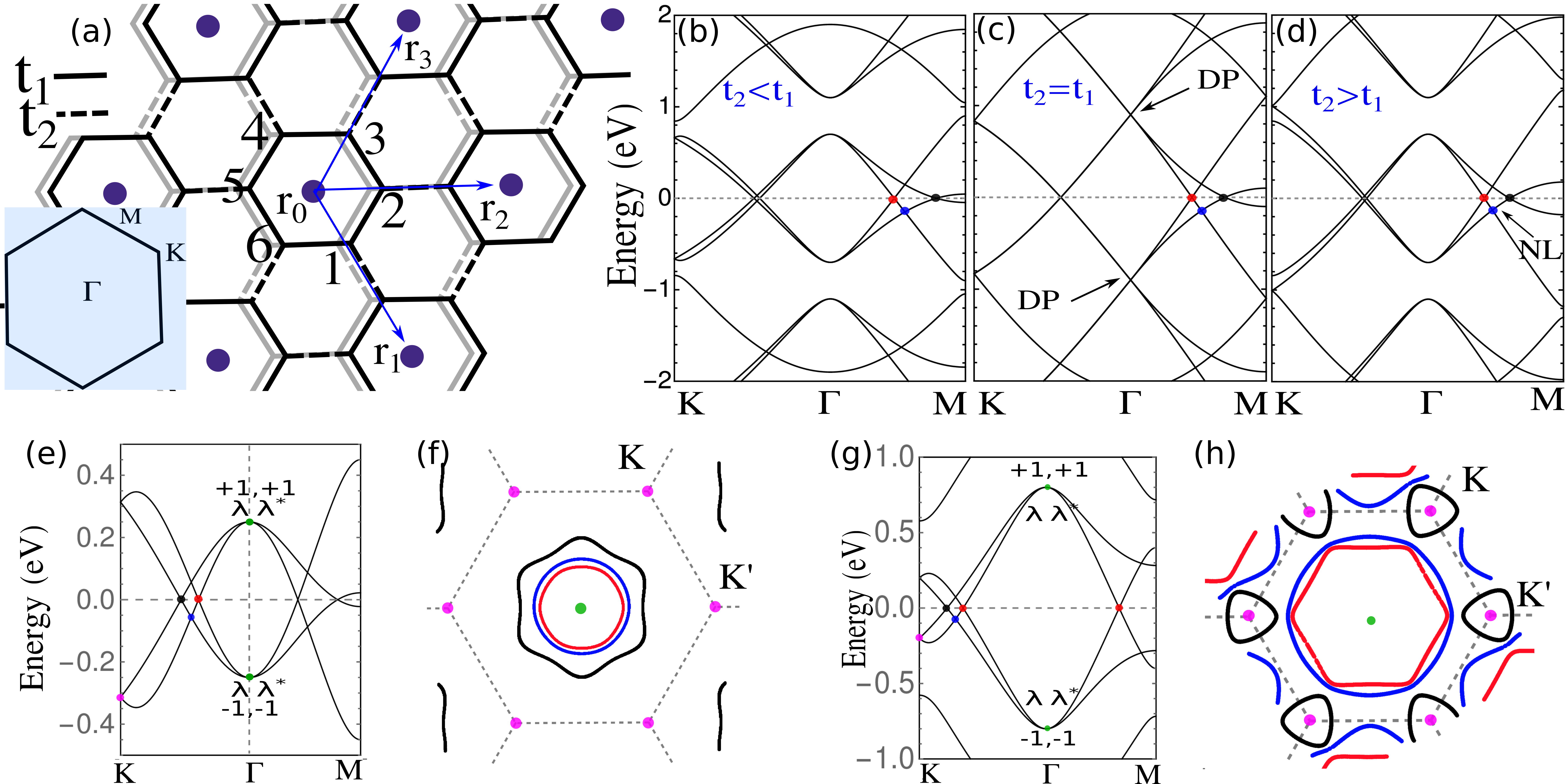}
	\caption{ 	
		(a) Structure of Kekul\'{e} textured AA-stacked bi-layer graphene (KABLG). 
		$t_1$ ($t_2$) indicates the intracell (intercell) hopping represented by solid (dashed) bonds. 
		The relevant first Brillouin zone (BZ) is shown in the inset. 
		(b,c,d) The band structures of KABLG when (b) $t_2<t_1$ (anti-Kekul\'{e} texture), (c) $t_2=t_1$, and (g) $t_2>t_1$ (Kekul\'{e} texture). In all cases, the same interlayer coupling $|t_3|<|t_1|, |t_2|$ is used.  
		(e, f) The band and nodal structures of KABLG near $E_F$ for parameters $t_1 =1$, $t_2 = 0.6$ and $t_3 =0.8$.
		$\pm 1$ ($\lambda$, $\lambda^*$) denote the $\mathcal{O}_L$ ($C_{3z}$) eigenvalues.
		(g, h) The band and nodal structures near $E_F$ when $t_3$=1.2 after a Lifshiftz transition of the black NL through the BZ boundary. 
		In (e,f,g,h), the red and black NLs are at $E_{F}$ while the blue NL is below $E_{F}$. 
		The magenta (green) NPs are below $E_{F}$ at the momentum $\bf K$/$\bf K^\prime$ ($\Gamma$). 
	}
	\label{4band}
\end{figure}

\par {\it Kekul\'{e}-textured AA-stacked bi-layer graphene (KABLG).|} 
The proposed 2D linked nodal structure can be realized in AA-stacked bi-layer graphene (BLG) with Kekul\'{e}-O distortion~\cite{Likek1,Likek}
which indicates the bond modulation pattern shown in Fig.~\ref{4band}(a) in which the intracell hopping $t_1$ within a hexagonal unit cell is distinguished from the intercell hopping $t_2$.
When two graphene layers with Kekul\'{e}-O distortion are vertically stacked (AA-stacking), 
we obtain the KABLG which can be realized by inserting metal ions between graphene layers.

We consider the following $12\times12$ tight-binding Hamiltonian for KABLG,
$H_{KABLG} = \sigma_0 \otimes H_{KSLG} + \sigma_x \otimes H_c$, 
where the Pauli matrices $\sigma_{x,y,z}$ denote the layer degrees of freedom and $\sigma_0$ is the related identity matrix. 
$H_{KSLG}$ indicates the Hamiltonian for a Kekul\'{e}-textured single layer graphene (KSLG) given by
$H_{KSLG}=-\sum_{<i,j>}t_{ij}c_i^{\dag}c_j + h.c$,     
where $c_i$ denotes the electron annihilation operator at the $i$-th site,
and $t_{ij}=t_1$ ($t_{ij}=t_2$) for the intracell (intercell) hopping between nearest-neighboring sites.
$H_c=diag[t_3,t_3,t_3,t_3,t_3,t_3]$ describes the interlayer hopping with the amplitude $t_3$ for nearest neighboring atoms between layers.
$H_{KABLG}$ has $\mathcal{I}$, $\mathcal{T}$, $\mathcal{C}$, $\mathcal{C}^{\prime}$ symmetries
represented by $\mathcal{I}=\sigma_x\otimes\tau_x\otimes I_3$,
$\mathcal{T}=K\sigma_0\otimes\tau_0\otimes I_3$, $\mathcal{C}=\sigma_z\otimes\tau_z\otimes I_3$,
$\mathcal{C}^{\prime}=\sigma_y\otimes\tau_z\otimes I_3$ where $I_3$ is a $3\times3$ identity matrix
related to the trimerization of the graphene unitcell induced by Kekul\'{e}-O distortion.
The Pauli matrices $\tau_{x,y,z}$ denote the sulattice degrees of freedom and $\tau_0$ is the related identity matrix.

Depending on the type of metal ions, the textured lattice can be in a phase with $t_2<t_1$ (anti-Kekul\'{e} distortion) or $t_2>t_1$ (Kekul\'{e} distortion) as shown in Fig.~\ref{4band}(b-d). 
In both cases, the nonzero interlayer hopping ($t_3$) pushes the gapped band structure of each textured graphene layer, upward and downward in energy, respectively, which generates linked NLs similar to those in Fig.~\ref{DBI}(e).

The band and nodal structures of the 4 bands near $E_F$ are similar in both $t_2<t_1$ and $t_1<t_2$ cases.
For convenience, here we focus on the $t_2<t_1$ case in which the 4 bands near $E_F$ are decoupled from other bands as shown in Fig.~\ref{4band}(b).
The $t_2>t_1$ case is discussed in the Supplemental Materials (SM)\cite{supply}.
As shown in Fig.~\ref{4band} (e, f), the band and nodal structures of the 4 bands near $E_F$ are identical to
those in Fig.~\ref{model} (i,j).
Namely, the two concentric NLs at $E_F$ are DC and linked with another NL below $E_F$.
As the two NLs at $E_F$ carry the same $Z^{0D}$ charge, they cannot be pair-annihilated through merging.

As $t_3$ increases, the size of the outer NL at $E_F$ becomes larger, and after touching the BZ boundary, 
it splits into two NLs encircling $K$ and $K^{\prime}$, respectively.
Interestingly, each of this NL develops another type of the linked nodal structure with a NP below $E_F$~\cite{L2}.
As each NL at $K$ or $K^{\prime}$ is DC, it cannot be annihilated separately (see SM\citep{supply}).
This is another way that the stability of two concentric NLs is manifested.

\begin{figure}[t!]
	\centering
	\includegraphics[width=0.5\textwidth]{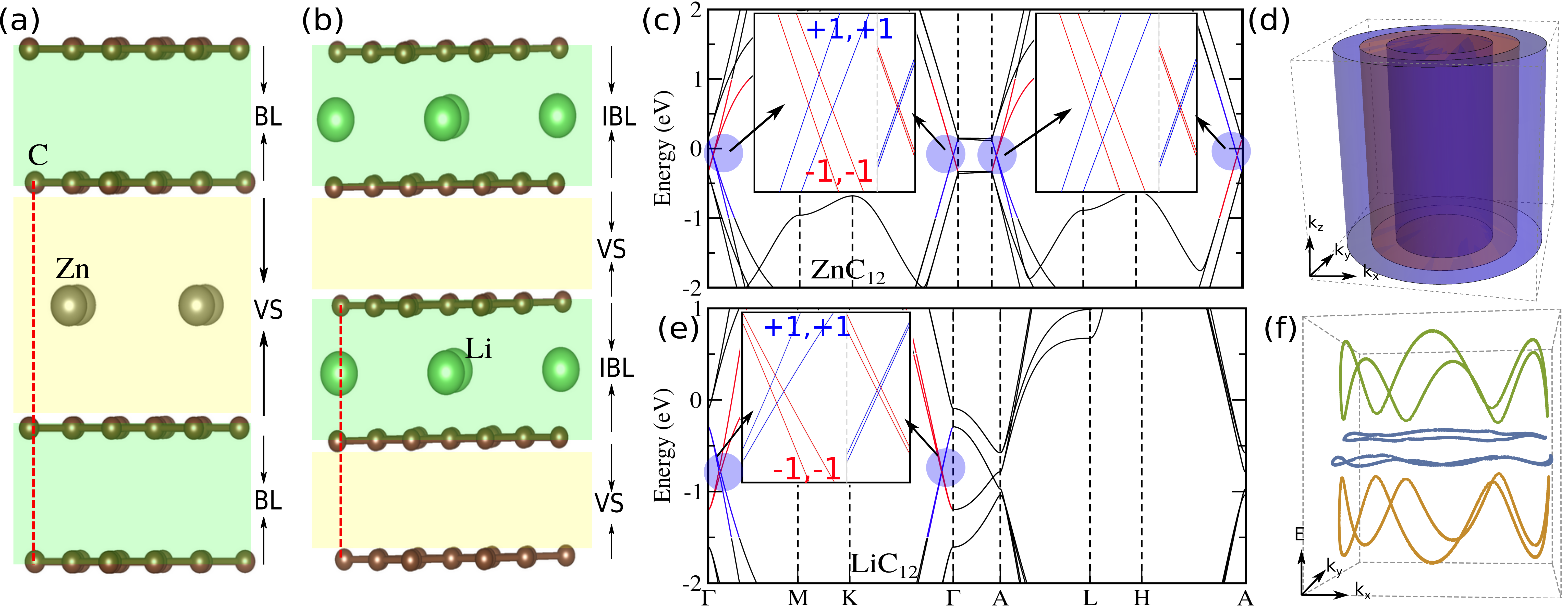}
	\caption{3D structures obtained by Van der Waals (VdW) stacking of BLGs.
		(a) Schematic structure of ZnC$_{12}$, a stacked BLG system with Zn atoms between neighboring BLGs. 
		Weak (repulsive) interaction between Zn atoms and BLG induces Kekul\'{e}-O type distortion of BLGs.  Here BL (VS) indicates the bilayer (VdW spacing). 
		(b) Schematic structure of LiC$_{12}$ where strong interaction between Li$^+$ ions and neighboring graphene layers gives intercalated bilayers (IBLs). 
		The red dashed vertical lines in (a,b) represent the unit cell length along the z-direction. 
		(c, e) Bulk band structures of (c) ZnC$_{12}$ and (e) LiC$_{12}$. Blue shaded regions represent the nodal degenerecies with close-up views in the inset. 
		$\pm1$ represents $\mathcal{O}_L$ ($M_z$) eigenvalues in (c) ((e)).
		(d) Schematic plot of the cylindrical concentric NSs at $E_F$ (blue) of ZnC$_{12}$ linked with another NS below $E_F$ (red). 
		(f) NLs of LiC$_{12}$ on the $k_z=0$ plane protected by $\mathcal{IT}$ and $M_z$ symmetries.
	}
	\label{DFT}
\end{figure}

\par {\it Candidate materials.|} 
We propose a class of 3D materials with chemical formula MC$_{12}$ (M: a metal ion) as a candidate system 
that exhibits the linked nodal structure discussed above.
We note that although the discussion up to now has been focused on 2D bilayer systems,
the same idea can be applied to 3D systems obtained by vertical stacking of 2D bilayers.
The linked concentric NLs in 2D bilayer systems can be naturally extended to the linked concentric NSs,
as long as the coupling between neighboring bilayers are weak enough so that the system preserves
the same symmetries including $\mathcal{C}$ and $\mathcal{C}^{\prime}$.

MC$_{12}$ has a hexagonal crystal structure with space group P${6/mmm}$, composed of Van der Waals (VdW) stacking of BLGs with intercalated M ions. 
Specifically, in ZnC$_{12}$ shown in Fig.~\ref{DFT}(a), the chemical interaction between Zn and C atoms is relatively weak,  
which can be supported by the fact that the optimized bond distance between Zn and C atoms is 3.40 $\AA$ that is larger than the sum of their VdW interaction lengths 3.10 $\AA$.
Although chiral symmetry is not exact in any material, it is an excellent symmetry in ZnC$_{12}$ due to the weak interaction between BLG layers. 
In accordance with this, ZnC$_{12}$ effectively respects two chiral symmetries,
and thus the linked cylindrical NSs can be realized.

The electronic structures of ZnC$_{12}$ are shown in Fig.~\ref{DFT} (c). 
In ZnC$_{12}$ with $\mathcal{O}_L$ symmetry, a DBI appears between two pairs of bands with different $\mathcal{O}_L$ eigenvalues
along $\Gamma$-A direction. 
The DBI generates two NSs near $E_F$ and another NS below the $E_F$
in the momentum region between the two NSs at $E_F$,
which is schematically described in Fig.~\ref{DFT}(d).
Similar linked NSs can appear in a wide class of materials in the form of MC$_{12}$ 
such as AlC$_{12}$, BeC$_{12}$ in which a DBI occurs along the entire $k_z$ direction ($\Gamma$A) (see SM\citep{supply}). 

On the other hand, the situation changes dramatically in a related compound LiC$_{12}$. [See Fig.~\ref{DFT}(b).]
In LiC$_{12}$, the optimized distance between Li and C atoms is 2.35 $\AA$, which is much smaller than the sum of their VdW interaction lengths 3.52 $\AA$. 
This is because of the ionic nature of Li atoms which strongly interact with the carbon p$_z$ orbitals in graphene. 
Thus, in contrast to ZnC$_{12}$, LiC$_{12}$ allows the second nearest-neighbor hopping between the layers mediated by Li ions, 
which breaks both chiral symmetries.
Thus, the NSs, which exist when chiral symmetries present, are mostly gapped except in the $k_z=0$ plane.
However, as $\mathcal{IT}$ still is a symmetry of the system,
the compounds fall into the AZ+$\mathcal{I}$ class AI in which NLs can appear~\cite{L1}.
Also, as $\mathcal{IT}$ symmetry supports the quantized Berry phase for these NLs,
flat drumhead surface states can appear on the boundary.
The NLs in LiC$_{12}$ are shown in Fig.~\ref{DFT} (f) and the related surface states are described in SM\citep{supply}.

\par {\it Discussions.|} 
Finally, let us discuss the experimental feasibility of our proposed model and related materials. 
The main ingredient of our model Hamiltonian is Kekul\'{e}-O order which can be realized in BLG-like systems with proper intercalation of atoms such as Li,Ca,Zn, etc. 
Such Kekul\'{e}-O ordered BLG systems have been synthesized and studied using angle-resolved photo-emission spectroscopy (ARPES) and transport measurements~\cite{Likek1, Likek2, Likek3, Cakek1,Cakek2, kek-NP}. 
For instance, the gap opening at the Dirac cone due to chiral symmetry breaking was observed in LiC$_{12}$ from ARPES measurement~\cite{Likek1}. 
Similarly, CaC$_{12}$ was also synthesized experimentally to study superconductivity~\cite{Cakek1,Cakek2}. 
Moreover, the chemical stability of ZnC$_{12}$ was theoretically shown by Srimanta \etal~\cite{TM-C} in which the binding energies of 
various transition metal intercalated BLG were computed~\cite{TM-C}.
Various other types of atoms such as Yb~\cite{Yb-Eu}, Eu~\cite{Yb-Eu}, Ge~\cite{Ge} have been intercalated in similar BLG structures, and their electronic structures have been explored using ARPES. 
All these works strongly support the possible experimental realization of our predicted linked nodal structures in future experiments.

\begin{acknowledgements}
C.M. thanks Rasoul Ghadimi for fruitful discussions.
C.M., S. K. and B.J.Y. were supported by the Institute for Basic Science in Korea (Grant No. IBS-R009-D1),
Samsung  Science and Technology Foundation under Project Number SSTF-BA2002-06,
the National Research Foundation of Korea (NRF) grant funded by the Korea government (MSIT) (No.2021R1A2C4002773, and No. NRF-2021R1A5A1032996).
\end{acknowledgements}




\end{document}


\beginsupplement


\title{Supplemental Material for \\ `` Unremovable linked nodal structures protected by crystalline symmetries in stacked bilayer graphene with Kekul\'{e} texture ''}

\author{Chiranjit Mondal}

\author{Sunje Kim}

\author{Bohm-Jung Yang}
\email{bjyang@snu.ac.kr}

\affiliation{Center for Correlated Electron Systems, Institute for Basic Science (IBS), Seoul 08826, Korea}
\affiliation{Department of Physics and Astronomy, Seoul National University, Seoul 08826, Korea}
\affiliation{Center for Theoretical Physics (CTP), Seoul National University, Seoul 08826, Korea}

\date{\today}

\maketitle 

\twocolumngrid

\tableofcontents

\newpage

This supplement contains details of tight binding (TB) calculations of the AA-stacked Kekul\'{e}-O texture bilayer graphene (KABLG) and MC$_{12}$ (M: a metal ion) systems, definition of various topological charges for the relevant nodal structures, Linking structures, and DFT computational details including the electronic structure of additional compounds.
 In section S1, we explicitly derive the TB Hamiltonian for KABLG and MC$_{12}$. In section S2. we derive the four bands continuum Hamiltonian for KABLG.
Section S3 is dedicated to the formal definition of topological charges for the relevant nodal structures. 
Here, we first review the nodal structure and topological charges of BDI class belonging to inversion symmetry $\mathcal{AZ+I}$ classification \cite{DCN3}. 
Then we discuss about the new nodal structure and corresponding topological charge when an additional chiral symmetry and $\mathcal{C}_3$ rotation symmetries are supplemented within the class BDI.
In Sec S4, we discuss the concept of Linking structure and apply it on AA-stacked bilayer graphene (BLG) and KABLG. 
Finally, in section S5, we have presented DFT calculation details and electronic structures of the additional materials for our proposed theory.

\begin{figure*}[t!]
\centering
\includegraphics[width=\linewidth]{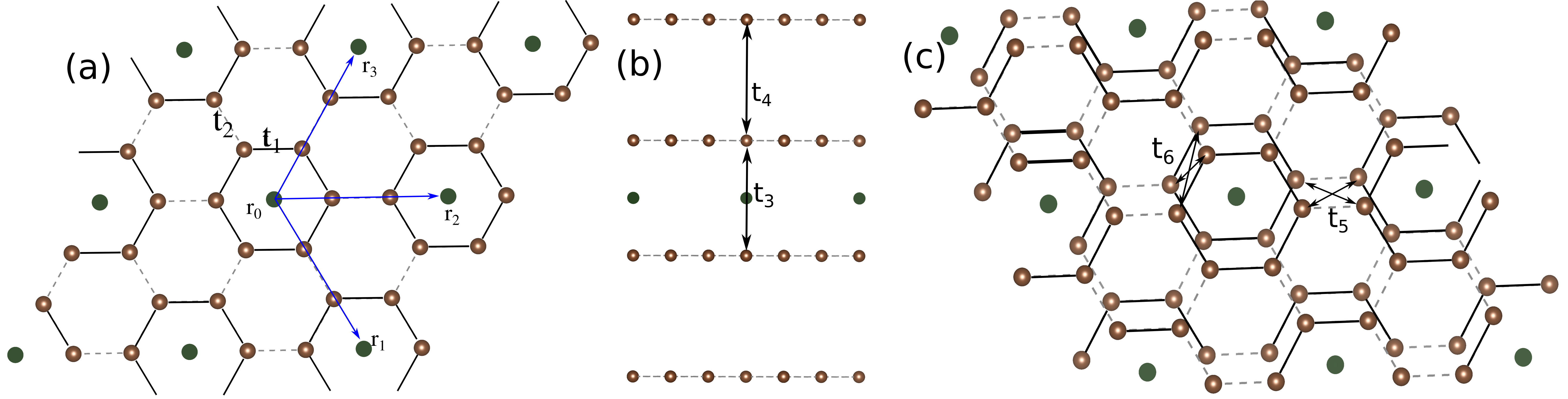}
\caption{(a) Single  layer graphene with Kekul\'{e}-O modulation with superlattice period ($\sqrt{3}\times\sqrt{3})$ in $x-y$ plane. $t_1$ and $t_2$ represents strong and weak bond hoppings. $r_1$, $r_2$ and $r_3$ are the vectors connecting to the nearest neighbor cells in the lattice. (b) Layer view of the bilayer structure. $t_3$ represents the interlayer hopping and $t_4$ counts the interaction between bilayers along k$_z$ direction. (c) $t_4$ and $t_5$ are the chiral symmetry breaking hopping terms.}
\label{S1}
\end{figure*}

\section*{S1. Tight binding calculation and symmetry of the Hamiltonian }
The Hamiltonian for KABLG can be written as,
\begin{equation}\label{blg}
H_{KBLG} = \sigma_0 \otimes H_{KSLG} + \sigma_x \otimes H_c
\end{equation}
where the Pauli matrices $\sigma_{x,y,z}$ denote the layer degrees of freedom and $\sigma_0$ is the related identity matrix and $H_{KSLG}$ represents the Hamiltonian for single layer Kekul\'{e} graphene (KSLG) which is given by
\begin{equation}\label{SLG}
H_{KSLG}=-t_1 \sum_{i=1}^{6} c^{\dagger}_{r_{0,i}} c_{r_{0,i+1}}
-t_2 \sum_{i=1}^{3} c^{\dagger}_{r_{0,i}} c_{r_{i,i+3}} + h.c,     
\end{equation}
where $c_i$ indicates the Fermionic annihilation operator of an electron at the $i$-th site. Intracell and intercell hoppings are represented by $t_1$ and $t_2$. 
$H_c$ represents the coupling Hamiltonian between the layers which is given by $H_c=diag[t_3,t_3,t_3,t_3,t_3,t_3]$. 
$t_3$ is the nearest neighbor (NN) coupling between the two layers.

 The vectors $r_1$,$r_2$ and $r_3$ is given by $r_1 = \frac{3a}{2}\hat{x}-\frac{3\sqrt{3}a}{2}\hat{y}$, $r_2 = 3a\hat{x}$ and $r_3 = \frac{3a}{2}\hat{x}+\frac{3\sqrt{3}a}{2}\hat{y}$. 
 To get the spectrum, we write the Hamiltonian H$_{KSLG}$ by taking the Fourier transform of the electron operators as, $c^{\dagger}_i=\frac{1}{\sqrt{N/2}} \sum_k e^{ik.r_i} c^{\dagger}_\mathbf{k}$. 
We express the electron wave functions for the Hamiltonian in six component spinors as
\begin{equation}
\Psi_{\mathbf{k}}=(c_{1\mathbf{k}},c_{2\mathbf{k}},c_{3\mathbf{k}},c_{4\mathbf{k}},c_{5\mathbf{k}},c_{6\mathbf{k}})^T.
\end{equation}
The Hamiltonian H$_{KSLG}$ =$\sum_{\mathbf{k}} \Psi_{\mathbf{k}}^{\dagger} H_{KSLG}(\mathbf{k}) \Psi_{\mathbf{k}}$. Where
\begin{equation}
\begin{aligned}
&  H_{KSLG}(\mathbf{k})= \\
&
 \begin{pmatrix}
0  & t_1 &  0     & t_2 e^{i \bf{k.r_1}}     & 0   & t_1    \\

t_1   & 0 & t_1  & 0     & t_2 e^{-i \bf {k.r_3}}   & 0     \\

0   & t_1  & 0 & t_1     & 0   & t_2 e^{-i \bf {k.r_2}}      \\

t_2 e^{-i \bf {k.r_1}}   & 0   & t_1   & 0   & t_1   & 0      \\

0   & t_2 e^{i \bf {k.r_3}}   & 0   & t_1     & 0 &  t_1      \\

t_1   & 0   & t_2 e^{i \bf {k.r_2}}   & 0     & t_1   & 0      \\
\end{pmatrix}
\end{aligned}
\end{equation}

The KABLG Hamiltonian takes form as $H_{KBLG}(\mathbf{k}) = \sigma_0 \otimes H_{KSLG}(\mathbf{k}) + \sigma_x \otimes H_c$.
$H_{KABLG}(k)$ has inversion ($\mathcal{I}$), time reversal ($\mathcal{T}$), chiral ($\mathcal{C}$), additional chiral ($\mathcal{C}^{\prime}$) symmetries which are represented by
\begin{equation}
\begin{aligned}
   \mathcal{I}=\sigma_x\otimes\tau_x\otimes I_3,           \\
   \mathcal{T}=K\sigma_0\otimes\tau_0\otimes I_3,           \\
   \mathcal{C}=\sigma_z\otimes\tau_z\otimes I_3,             \\
   \mathcal{C}^{\prime}=\sigma_y\otimes\tau_z\otimes I_3 
\end{aligned}
\end{equation}
 where $I_3$ is a $3\times3$ identity matrix related to the trimerization of the graphene unitcell induced by Kekul\'{e}-O distortion.
The Pauli matrices $\tau_{x,y,z}$ denote the sublattice degrees of freedom and $\tau_0$ is the related identity matrix.

\begin{figure*}[t!]
\centering
\includegraphics[width=\linewidth]{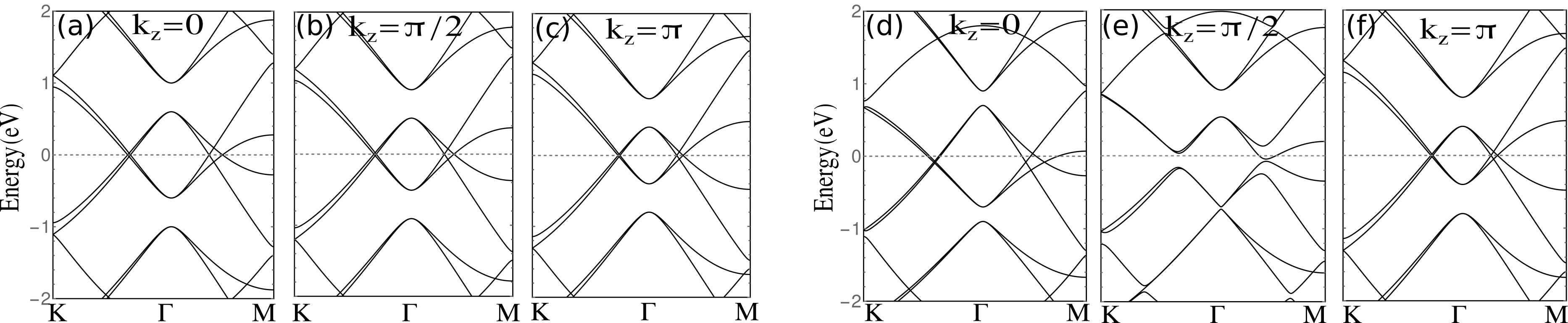}
\caption{Band structures are calculated from tight binding Hamiltonian with chiral symmetry (a-c) using parameters $t_1=1$, $t_2=0.9$, $t_3=0.7$ and $t_4=0.1$ and without chiral symmetry (d-f) using parameters $t_1=1$, $t_2=0.9$, $t_3=0.7$, $t_4=0.1$, $t_5 = 0.1$, and $t_6=0.1$.} 
\label{S2}
\end{figure*}

We can construct AC$_{12}$ (A=Zn,Al,Be) lattice by vertically stacking the KABLG along z-direction. 
Corresponding Hamiltonian can be constructed by adding a term $I_6 \otimes t_4 e^{ik_z}$ with the coupling Hamiltonian $H_c$ in Eq.~\ref{blg} where $I_6$ is a $6\times 6$ identity matrix and $t_4$ represents the hoppings between the NN of the two KABLG.
The energy dispersion of the Hamiltonian gives concentric nodal cylinder about the $k_z$ axis, as long as the coupling between
neighboring bilayers are weak enough so that the system preserves the same symmetries of KABLG including $\mathcal{C}$ and $\mathcal{C}^\prime$.
The product of these two chiral ($i \mathcal{C}\mathcal{C}^\prime$) gives a conserved local in $\bf k$
symmetry $\mathcal{O}_L$ (i.e, $\mathcal{O}_L H_{KBLG}({\bf k}) \mathcal{O}^{-1}_L = H_{KBLG}({\bf k})$) which stabilized the nodal structure when two bands cross each other with opposite eigen values of $\mathcal{O}_L$.
In KABLG,  $\mathcal{O}_L = \sigma_x \otimes \sigma_0 \otimes I_3 $, which physically represents the layer exchange symmetry between  the layers in KABLG. 
The band structures of MC$_{12}$ (M=Zn,Al,Be) are shown in Fig.~\ref{S2}(a-c) for different values of $k_z$.

We then the construct the Hamiltonian for BC$_{12}$(B=Li,Mg,B) by further adding chiral symmetry breaking hoppings $t_5$  and $t_6$ as shown in Fig.~\ref{S1}(c).
 In contrast to AC$_{12}$ (A=Zn,Al,Be), this type of hoppings are not negligible in case of MC$_{12}$ BC$_{12}$ (B=Li,Mg,B).
  These hoppings $t_5$  and $t_6$ break both the chiral symmetries of the Hamiltonian which open up gaps at the nodal cylinders except $k_z=0$ and $k_z=\pi$ planes protected by $\mathcal{PT}$ and $\mathcal{M}_z$ symmetries. 
  The Hamiltonian represents a nodal line (NL) structure.
The chiral symmetry braking term can be written as,
$ H_{56}(\bf k)=
 \begin{pmatrix}
0  & t_{56}({\bf k})     \\
t^{\dagger}_{56}({\bf k})   & 0    \\
\end{pmatrix}
$
where $t_{56}$ is given by

\begin{equation}
t_{56}(\bf k)=
 \begin{pmatrix} 
 0   &   t_5  &  0  &   x_1  &   0     &  t_5   \\
 t_5   &   0  &  t_5  &   0    &   y_1   &  0   \\
 0   &   t_5  &  0  &   t_5    &   0     &  z_1   \\
 x_2 &   0  &  t_5  &   0    &   t_5    &  0   \\
 0  &  y_2   & 0  &   t_5  &  0   &  t_5   \\
 t_5   &   0  & z_2 &   0    &   t_5     &  0   \\
\end{pmatrix}
\end{equation}

and

\begin{equation}
\begin{aligned}
&  x_1 = t_6 Exp[i(\frac{3ak_x}{2} - \frac{3\sqrt{3}ak_y}{2} + k_z)]      \\
&  y_1 = t_6 Exp[-i(\frac{3ak_x}{2} + \frac{3\sqrt{3}ak_y}{2} - k_z)]   \\
&  z_1 = t_6 Exp[-i(3k_x a - k_z)]                                        \\
&  x_2 = t_6 Exp[-i(\frac{3ak_x}{2} - \frac{3\sqrt{3}ak_y}{2} - k_z)]      \\
&  y_2 = t_6 Exp[i(\frac{3ak_x}{2} + \frac{3\sqrt{3}ak_y}{2} + k_z)]        \\
&  z_2 = t_6 Exp[i(3k_x a + k_z)]
\end{aligned}
\end{equation}

 The electronic structures for the the Hamiltonian without chiral symmetry (which represents the band structure of LiC$_{12}$, BC$_{12}$, MgC$_{12}$) are shown in Fig.~\ref{S2}(d-f).  The nodal degeneracies are protected only in $k_z=0$ and $k_z=\pi$ planes.

\subsection*{1. Projection of KABLG TB Hamiltonian on 4 low energy bands }
We project the $12 \times 12$ tight binding KABLG Hamiltonian on the four low energy bands which effectively take part in the nodal structure.
First, we compute the eigen function of the KABLG TB Hamiltonian (Eq.~\ref{blg}) for $t_1 = 1, t_2=0.6$ and $t_3=0.8$ at $\Gamma$ point. Then we project the $12 \times 12$ Hamiltonian on the four low energy bands around the $E_F$. Expanding it up to second order of $\bf k$, we obtain the $4 \times 4$ low energy Hamiltonian as, 
\begin{equation}\label{proj}
\begin{aligned}
H_{4 \times 4} =
& (-0.4 + 1.35~k_x^2 +1.35~k_y^2)~\sigma_z \tau_0 -1.35~k_x k_y ~ \sigma_z \tau_x   \\
& + (0.67~k_x^2  - 0.67~k_y^2)~\sigma_z \tau_z 
\end{aligned}
\end{equation}

Two chiral, Inversion and time reversal symmetries are present in the above Hamiltonian which are given by,
\begin{equation}
\begin{aligned}
\mathcal{C}= \sigma_x \tau_0, \\
\mathcal{C}^\prime = \sigma_y \tau_0, \\
\mathcal{I}= \sigma_0 \tau_0, \\
\mathcal{T}= K \sigma_0 \tau_0, \\
\end{aligned}
\end{equation}
     
Please note that the representation of the symmetries apparently changes when projecting the $12 \times 12$ bands TB Hamiltonian on the four low energy bands around the $\Gamma$ point. It is because we lose some degrees of freedom in the process. However, as the projection operator always has an ambiguity, we can choose suitable unitary operation ($\mathcal{U}_r$) to rotate the basis to see the sameness of the representation of the $12 \times 12$ bands TB Hamiltonian and projected $4 \times 4$ low energy Hamiltonian. For instance, choosing $\mathcal{U}_r$ as $\mathcal{U}_\sigma \otimes \tau_0$ where $\mathcal{U}_\sigma$ works on the layers and given by
\begin{equation}
\begin{aligned}
& \mathcal{U}_\sigma = \frac{1}{\sqrt{2}}
 \begin{pmatrix}
1     &      -i     \\

1     &       i     \\
\end{pmatrix}
\end{aligned}
\end{equation}
and $\tau_0$ is $2 \times 2$ identity matrix, the basis can be rotated as 
$ \sigma_x \mapsto \sigma_y \mapsto \sigma_z \mapsto \sigma_x$. In the rotating basis, representation of projected low energy model becomes
\begin{equation}
\begin{aligned}
\mathcal{C}= \sigma_y \tau_0, \\
\mathcal{C}^\prime = \sigma_z \tau_0, \\
\mathcal{I}= \sigma_0 \tau_0, \\
\mathcal{T}= K \sigma_0 \tau_0, \\
\end{aligned}
\end{equation}
   
Even though some degrees of freedom become obsolete after low energy projection, one can see the some similarity between the symmetry representation of the TB and projected low energy Hamiltonian after rotating the basis. For instance, the layer exchange symmetry ($\mathcal{O}_L = i \mathcal{C}\mathcal{C}^\prime$) is represented by $\sigma_x$ in both model.

\section*{S2. General four bands Hamiltonian in class BDI with inversion symmetry }
A system belonging to the $\mathcal{AZ}$ class BDI respects three  symmetries, which are time-reversal symmetry ($\mathcal{T}$), chiral symmetry($\mathcal{C}$), particle-hole symmetry ($\mathcal{P}$) \cite{DCN3,AZ}.
In the presence of an additional inversion symmetry ($\mathcal{I}$), a local in $\bf k$ antiunitary constraints is imposed on $\mathcal{T}$ and $\mathcal{P}$. Apart from the non-local nature in $\bf k$ space, the combine symmetries $\mathcal{TI}$ and $\mathcal{PI}$ fulfill the similar symmetry constraints like $\mathcal{T}$ and $\mathcal{P}$ which are given by 
\begin{equation}
\begin{aligned}
\mathcal{(TI)} \mathcal{H(\bf k)}  \mathcal{(TI)}^{-1} = \mathcal{H(\bf k)} , \\
\mathcal{(PI)} \mathcal{H(\bf k)}  \mathcal{(PI)}^{-1} = -\mathcal{H(\bf k)} , \\
\mathcal{C} \mathcal{H(\bf k)}  \mathcal{C}^{-1} = -\mathcal{H(\bf k)} . \\
\end{aligned}
\end{equation}
We also impose the $\mathcal{I}$ symmetry as the double band inversion occurs at the $\Gamma$ point in KABLG.

 To find out a matrix representation of these symmetries, we have to consider some conditions. 
 First, $\mathcal{(TI)}^2=\mathcal{(PI)}^2=1$ since KABLG belongs to the inversion symmetric class BDI. 
 Second, $\mathcal{I}$ is commute with $\mathcal{T}$ since $\mathcal{I}$ is a spatial symmetry.
  Also, $\mathcal{C}$ of KABLG comes from the sublattice symmetry. This means $\mathcal{C}$ is a spatial symmetry and it is commute with $\mathcal{T}$.
   From the above conditions, we can choose a matrix representation which is given by
\begin{equation}
\begin{aligned}
\mathcal{T}=\sigma_0 \tau_0 K,\\
\mathcal{C}=\sigma_x \tau_0,\\
\mathcal{I}=\sigma_0 \tau_0,\\
\mathcal{P}=\mathcal{TC}=\sigma_x \tau_0 K,
\end{aligned}
\end{equation}
where $K$ is the complex conjugation operator.

The general 4-band Hamiltonian is given by the linear combination of $\sigma_i \tau_j$ matrices with $(i,j)=0,x,y,z$. 
But some of them should be removed due to the symmetries. 
To check which terms will be survived, let us consider the symmetry constraints arising from the local symmetries in the momentum space.
 There are two independent local symmetries in class BDI$+\mathcal{I}$, which are $\mathcal{TI}, \mathcal{C}$. 
 Since $\mathcal{TI}=K$, only 10 real-valued matrices among $\sigma_i \tau_j$ will be survived. Those matrices are given by
\begin{gather}
\sigma_0 \tau_0,~\sigma_0 \tau_x,~\sigma_0 \tau_z,           \nonumber   \\ 
\sigma_x \tau_0,~\sigma_x \tau_x,~\sigma_x \tau_z,            \nonumber   \\
\sigma_y \tau_y,~\sigma_z \tau_0,~\sigma_z \tau_x,            \nonumber    \\
\sigma_z \tau_z.                                             \nonumber
\end{gather}
Also matrices commuting with $\mathcal{C}$ should be removed. Then the remaining matrices are given by
\begin{gather}
\sigma_y \tau_y,~\sigma_z \tau_0,~\sigma_z \tau_x,~\sigma_z \tau_z. \nonumber
\label{basis}
\end{gather}
So the general Hamiltonian is given by 
\begin{gather}
H(\mathbf{k})=p(\mathbf{k})\sigma_z \tau_0+q(\mathbf{k}) \sigma_z \tau_x +r(\mathbf{k}) \sigma_z \tau_z +s(\mathbf{k})\sigma_y \tau_y .
\label{genH}
\end{gather}

The coefficients of each matrices are also affected by a non-local symmetry $\mathcal{I}$. Since $\mathcal{I}=\sigma_0 \tau_0$, all the coefficients in Eq. (\ref{genH}) are even function of $\mathbf{k}$.

In KABLG, there is one more chiral symmetry $\mathcal{C}^\prime = \sigma_y \tau_0$. This put an additional condition that is $s(\mathbf{k})=0$.

\subsection*{1. Double band inversion of KABLG}
Let us consider a Hamiltonian, Eq. (\ref{genH}), where the coefficients are given by
\begin{gather}
p(\mathbf{k})=M,~q(\mathbf{k})=k^2-m,~r(\mathbf{k})=s(\mathbf{k)}=0,
\label{coeff}
\end{gather}
where $M,m$ are constants with a condition $M>0$. Increasing $m$, we can see the double band inversion process from the Hamiltonian with coefficients Eq. (\ref{coeff}).

While $m$ is smaller then $-M$, there is no nodal lines. When $m$ becomes bigger than $-M$, a nodal line at $E_F$ appears near $\mathbf{k}=0$. When $m$ becomes positive, a nodal line below $E_F$ appears near $\mathbf{k}=0$. When $m$ becomes bigger than $M$, another nodal line at $E_F$ appears near $\mathbf{k}=0$. As a result, we can get linked nodal lines near $\mathbf{k}=0$ from double band inversion process. In the main paper, the process is illustrated.

To find a robust linking structure, we need to consider a $C_3$ symmetry like KABLG. Let us impose a matrix representation $C_3=e^{I \frac{2\pi}{3} \sigma_0 \tau_y}$, which satisfies $(C_3)^3=1,~[C_3,T]=0$. Then the $C_3$ symmetry relation is given by
\begin{gather}
(C_3)^{-1} H(\mathbf{k}) C_3 = H(\mathbf{k'}),
\end{gather}
where ${k'}_x=-\frac{1}{2}k_x -\frac{\sqrt{3}}{2}k_y ,~{k'}_y=\frac{\sqrt{3}}{2}k_x -\frac{1}{2}k_y$. To impose this symmetry, the coefficients $p(\mathbf{k}),q(\mathbf{k}),r(\mathbf{k})$ in the Hamiltonian should be changed. Let us define new $C_3$-symmetric coefficients by
\begin{gather}
p(\mathbf{k})=M+2\alpha (k_{x}^2+k_{y}^{2}),\label{coe1}\\
q(\mathbf{k})=-2\alpha k_x k_y\label{coe2} \\
r(\mathbf{k})=\alpha (k_{x}^{2}-k_{y}^{2}).\label{coe3}
\end{gather}
Note that constant terms of $q(\mathbf{k})$ and $r(\mathbf{k})$ are zero due to the $C_3$ symmetry.

From the coefficients (\ref{coe1}), (\ref{coe2}), (\ref{coe3}), we can see that the robust linking structure arises from a double band inversion process. Let us suppose $\alpha>0$ and change $M$. The eigenvalues of the Hamiltonian is given by
\begin{gather}
M+3\alpha k^2 ,\label{band4}\\
M+\alpha k^2 ,\label{band3}\\
-(M+\alpha k^2) ,\label{band2}\\
-(M+3\alpha k^2) \label{band1} .
\end{gather}

Note that bands Eq. (\ref{band4}) and Eq. (\ref{band3}) have a degenerate point at $\Gamma$. This is because they have complex $C_3$ eigenvalues at $\Gamma$. This is the same for the bands Eq. (\ref{band2}) and Eq. (\ref{band1}). Also, there is a layer-interchanging symmetry $\mathcal{O}_L = \sigma_z \tau_0$ and bands Eq. (\ref{band4}), Eq. (\ref{band3}) (Eq. (\ref{band2}), Eq. (\ref{band1})) have eigenvalues 1(-1) for $\mathcal{O}_L$.

When $M>0$, there is no nodal line at $E_F$. But $M$ becomes negative, two nodal lines appear at $E_F$ with radii $\sqrt{|M|/\alpha}$ and $\sqrt{|M|/3\alpha}$, respectively. Each nodal line is topologically protected due to the $\mathcal{O}_L$ symmetry eigenvalues. In the same time, a nodal line below $E_F$ appears with a radius $\sqrt{|M|/2\alpha}$. This nodal line is also topologically protected due to the same reason. Therefore, we can get a robust linking structure from this model. 
In the main paper, the process is illustrated.

\section*{S3. Topological charges }
Based on the intrinsic symmetries, the topological space of Hamiltonian was classified ($\mathcal{AZ}$-classification) \cite{AZ}. The classification scheme further extended to centro-symmetric systems ($\mathcal{AZ+I}$-classification) \cite{DCN3}. The Kekul\'{e} texture bilayer graphene belongs to BDI class of $\mathcal{AZ+I}$ classification. Here, we briefly discuss about the symmetries and topological charges of BDI class.

\subsection*{1. 0D and 1D $\mathbb{Z}_2$ charges}
In 2D (3D) system, BDI symmetries allow NLs (NSs) at $E_F$ which are linked with a NP (NL) below the $E_F$ \cite{L2}.
In general, the topological charge of NP (NL) below the $E_F$ is equivalent to Berry phase which can be computed using smooth and complex wavefunctions. Another way to compute this charge is to check the orientation of the wavefunctions using a real gauge (formally know as Stiefel-Whitney class\cite{L1}). The reality condition of the wavefunctions is enforced by the $\mathcal{PT}$ symmetry.  

The charges of the NL (NS) at $E_F$ can be defined using the notion of homotopy equivalence of the classifying space of the Hamiltonian.
 We calculate the 0D (1D) homotopy charges ($\pi_0$ and $\pi_1$) on the 0D (1D) manifolds ($S^0$ and $S^1$) enclosing the NL (NS) at $E_F$ in a 2D (3D) system. 
 The general strategy to compute the homotopy charges are as follows \cite{DCN1,DCN2};
  We consider a non-contactable $n^{th}$ sphere $S^n$ as the manifold enclosing the node. 
  All the possible $S^n\subset$ BZ with a gaped spectrum can be continuously mapped to a topological space $M$. 
  Here, $M$ represents the Hilbert space spanned by all occupied bands.
   Hamiltonian ($H(\bf k$)) represents the mapping function. 
   Continuous deformation of $S^n$ or $H(\bf k$) correspond to different mapping which can be classified by \textit{equivalence class}.
    All the mapping in \textit{equivalence class} are isomorphic to each other. 
    The elements of the homotopy group $\pi_n(M)$ are represented by the distinct \textit{equivalence class} of such mapping \cite{DCN1,DCN2}.
     For instance, if $\pi_n(M)$ contains only identity element (say, homotopy group of a sphere), all the mapping are trivial mapping. 
     However, if the homotopy group is $\mathbb{Z}_2$, two non-equivalent class correspond to trivial and non-trivial mapping. 

The NL (NS) at $E_F$ in class BDI may have two charges; 0D and 1D charge. 
The interpretation of these charges are as follows.
class BDI possess chiral symmetry.
 If we choose chiral symmetry $\mathcal{C}=\sigma_z$, the Hamiltonian takes the block off-diagonal form as 
\begin{equation}
 H(\mathbf{k})=
 \begin{pmatrix}
0 & A(\mathbf{k}) \\
A^T(\mathbf{k}) & 0\\
\end{pmatrix}
\end{equation}
Because of the additional $\mathcal{PT}$ symmetry, $A(\mathbf{k})$ can be represented by N$\times$N real matrix for a 2N bands Hamiltonian. The 0D charge for such Hamiltonian is defined as \cite{DCN3,L2}, 
\begin{equation}
c_{BDI}(S^0)= sgn\Big[\prod_{k\in S^0} det A(\mathbf{k})\Big]\in \{+1,-1\}
\end{equation}
where, $S^0=\{\mathbf{k}_{in},\mathbf{k}_{out}\}$, represent two point inside and outside the NL (NS).

 Because of the chiral symmetry, N occupied bands are related to N unoccupied bands. For instance, an occupied states $\ket{u^{occ}_{n\bf{k}}}$ with energy $-E_{n\bf{k}}$ possess corresponding chiral symmetric states $C \ket{u^{occ}_{n\bf{k}}}$ at energy $+E_{n\bf{k}}$. Lets express the eigen states of occupied and unoccupied states as

\begin{equation}
\begin{aligned}
\ket{u^{occ}_{n\bf{k}}}=\frac{1}{\sqrt{2}}
 \begin{pmatrix}
\ket{u^{\uparrow}_{n\bf{k}}}  \\
\ket{u^{\downarrow}_{n\bf{k}}}  \\
\end{pmatrix} \\
\ket{u^{unocc}_{n\bf{k}}}=\frac{1}{\sqrt{2}}
 \begin{pmatrix}
\ket{u^{\uparrow}_{n\bf{k}}}  \\
-\ket{u^{\downarrow}_{n\bf{k}}}  \\
\end{pmatrix}
\end{aligned}
\end{equation}

where $\ket{u^{\uparrow}_{n\bf{k}}}$ and $\ket{u^{\downarrow}_{n\bf{k}}}$ represent the N-dimensional vector and satisfy the following relations, 
\begin{equation}
\begin{aligned}
\braket{u^{\uparrow}_{n\bf{k}}}{u^{\uparrow}_{m\bf{k}}} = \braket{u^{\downarrow}_{n\bf{k}}}{u^{\downarrow}_{m\bf{k}}} = \delta_{nm}  \\
\mathlarger{\sum}_n \ket{u^{\uparrow}_{n\bf{k}}} \bra{u^{\uparrow}_{n\bf{k}}} = \mathlarger{\sum}_n \ket{u^{\downarrow}_{n\bf{k}}} \bra{u^{\downarrow}_{n\bf{k}}} = \mathds{1} .
\end{aligned}
\end{equation}

 The Hamiltonian becomes real because of the $\mathcal{PT}$ symmetry and the off-diagonal block can be express in terms of N-dimensional vectors outside the NL (NS) as 

\begin{equation}
A({\bf{k}}) = \mathlarger{\sum}_n E_{n\bf{k}} \ket{u^{\uparrow}_{n\bf{k}}} \bra{u^{\downarrow}_{n\bf{k}}}
\end{equation}
However, the relative orientation of the state vectors of the highest occupied and lowest unoccupied states change after the band inversion at NL (NS). 
For instance, suppose that highest occupied and lowest unoccupied are given by 
\begin{equation}
\begin{aligned}
\ket{u^{occ}_{1\bf{k}}}=\frac{1}{\sqrt{2}}
 \begin{pmatrix}
\ket{u^{\uparrow}_{1\bf{k}}}  \\
\ket{u^{\downarrow}_{1\bf{k}}}  \\
\end{pmatrix} \\
\ket{u^{unocc}_{1\bf{k}}}=\frac{1}{\sqrt{2}}
 \begin{pmatrix}
\ket{u^{\uparrow}_{1\bf{k}}}  \\
-\ket{u^{\downarrow}_{1\bf{k}}}  \\
\end{pmatrix}
\end{aligned}
\end{equation}
After band inversion, i,e inside the NL (NS), the off-diagonal block becomes, 
\begin{equation}
A(\textbf{k}) = -E_{n\bf{k}}\ket{u^{\uparrow}_{1\bf{k}}} \bra{u^{\downarrow}_{1\bf{k}}} + \mathlarger{\sum}_n E_{n\bf{k}} \ket{u^{\uparrow}_{n\bf{k}}} \bra{u^{\downarrow}_{n\bf{k}}}
\end{equation}
Therefor, the sign of the det $A(\mathbf{k})$ changes relative to the side of the NL (NS) because of the relative change of the orientation of the state vectors $\ket{u^{\uparrow}_{1\bf{k}}}$ and $\ket{u^{\downarrow}_{1\bf{k}}}$. 
Thus the sign of the $det A(\mathbf{k})$ defines the 0D $\pi_0$ charge for the NL (NS) at the $E_F$ in class BDI.

To calculate the 1D $\pi_1$ charge of the NL (NS) at E$_F$, we follow the spectral flattening technique. 
First we choose a $S^1$ surrounding the NL (NS) then decompose the Hamiltonian on the $S^1$ by its eigen system as 
 
\begin{equation}
H(\mathbf{k}) = \mathlarger{\sum}_{n=1}^{2N} E_{n\bf{k}} \ket{u_{n\bf{k}}} \bra{u_{n\bf{k}}}
\end{equation}

which can be deformed continuously to a flat-band Hamiltonian as 

\begin{equation} \label{FHAM}
H^{FB}(\mathbf{k}) = \mathlarger{\sum}_{n=1}^{2N} sign[E_{n\bf{k}}] \ket{u_{n\bf{k}}} \bra{u_{n\bf{k}}}.
\end{equation}

In class BDI, $A^{FB}(\mathbf{k}) \in O(N)$ on the $S^1$ where $A^{FB}(\mathbf{k})$ is the off-diagonal block of the flattened Hamiltonian $H^{FB}(\mathbf{k})$ in Eq.~\ref{FHAM}. 
 We restrict $A^{FB}(\mathbf{k})$ in certain $\mathbf{k}$ within the $SO(N)$ by taking the composition with mirror symmetry if $det A^{FB}(\mathbf{k})=-1$.
  Otherwise we keep it unchanged.
  The $\pi_1$ charge of the NL (NS) on $S^1$ is defined by the relation \cite{DCN3,DCN1,DCN2},

\begin{equation}
\pi_1(S^1) = [A(\mathbf{k}):S^1\rightarrow SO(N)] 
\end{equation} 
which means the homotopy equivalence of $SO(N)$ group.

\subsection*{2. 0D $\mathbb{Z}$ charge} 
When an additional chiral symmetry ($C^\prime$) is supplemented, we can define a $\bf k$ local symmetry $\mathcal{O}_L$. 
$Z^{0D}$ is defined by the eigenvalues ($\pm 1$) of $\mathcal{O}_L$ symmetry.
At a given momentum $\mathbf{k}$, we denote the number of occupied bands with the positive (negative) $\mathcal{O}_L$ eigenvalues by $\eta_+(\mathbf{k})$ ($\eta_-(\mathbf{k})$).
 Also we define 
 \begin{equation}
 \Delta\eta(\mathbf{k}) = \eta_+(\mathbf{k}) - \eta_-(\mathbf{k}).
\end{equation} 
For a given NL, we pick two momenta $\mathbf{k}_{in}$ and $\mathbf{k}_{out}$, which are inside and outside of the loop, respectively. 
Then we define the $Z^{0D}$ charge as
 \begin{equation}
 Z^{0D} \equiv \frac{1}{2} \left[\Delta\eta(\mathbf{k}_{\text{out}})-\Delta\eta(\mathbf{k}_{\text{in}})\right].
 \end{equation}

\section*{S4. Linking structure}
The linking structure is given by the following conditions \cite{L2,L1}. (i) Nodes at the $E_F$ have to have two distinct topological charges. (ii) The charge of the nodes at and below the E$_F$ have to be equivalent. (iii) The charge of nodes at E$_F$ is non-trivial  if a linking node is present below the E$_F$. 
In other words, if we trivialize the charge of the linking node below the $E_F$, the nodal structure at $E_F$ becomes unrobust.

\subsection*{1. 0D $\mathbb{Z}$ charge linking structure in KABLG}
For KABLG,  $\mathcal{C}$ impose a condition on the flattened Hamiltonian (Eq. \ref{FHAM}) that the off-diagonal block $A^{FB}(\mathbf{k}) \in O(N)$. $O(N)$ group have to be disconnected space, determined by the determinant $\pm 1$. This defines the $Z^{0D}$ $Z_2$ 0D charge charge of the NL (NS) at $E_F$. Please note that this definition is invalid for the nodes below the $E_F$ as the chiral symmetry only a good symmetry at $E_F$.

\begin{figure}[t!]
\centering
\includegraphics[width=\linewidth]{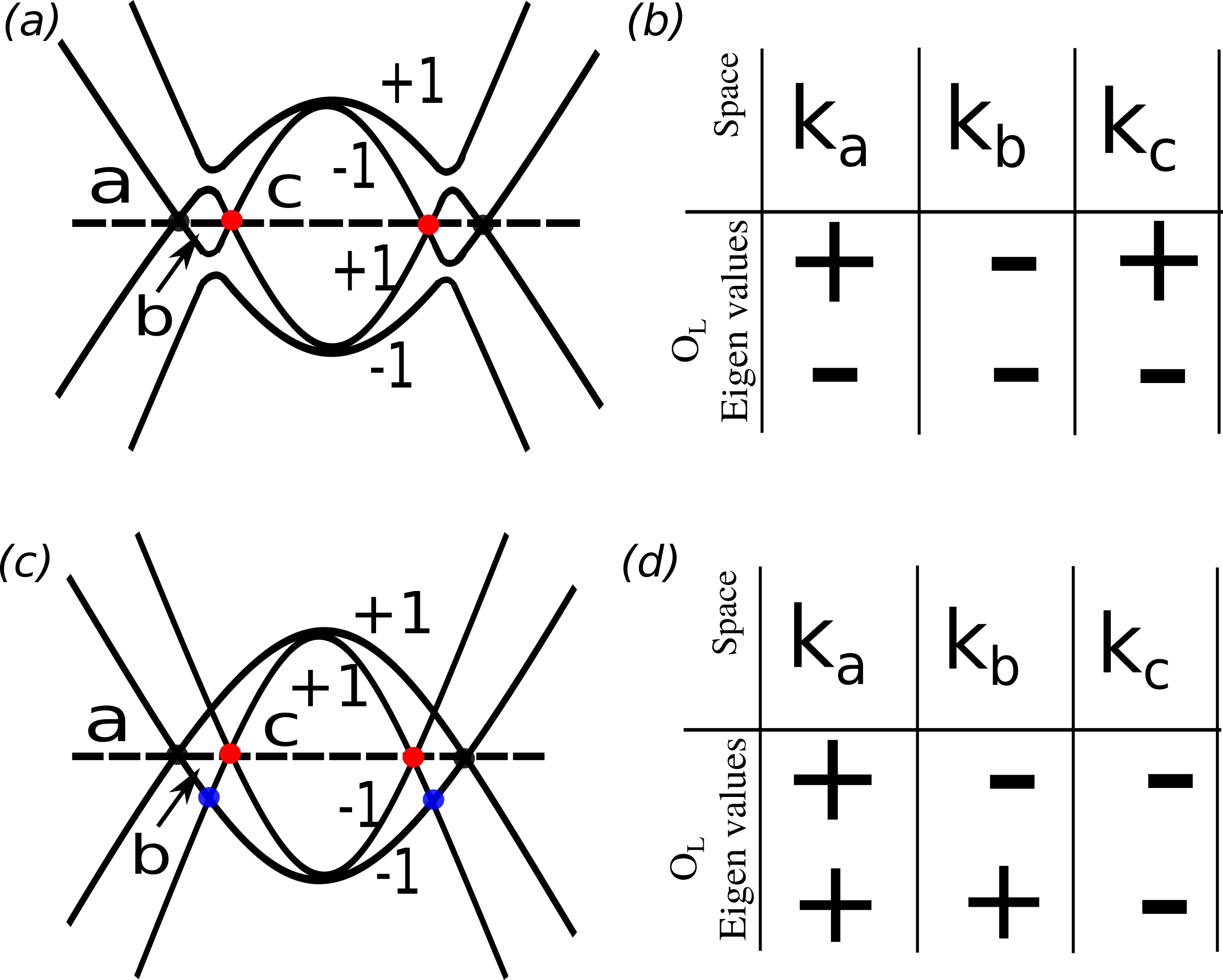}
\caption{(a,c) Band structure for two different configuration of $\mathcal{O}_L$ eigen values over the bands. (b,d) $\mathcal{O}_L$ eigen values are listed in certain ${\bf k}$ for the two cases.}
\label{Zc}
\end{figure}

It is easy to show that the other two conditions (ii and iii) are full filed naturally in DBI class when additional chiral symmetry ($\mathcal{O}_L$) and $\mathcal{C}_3$ rotation symmetries are supplemented. 
Both the NL (NS) at and below the $E_F$ have $Z^{0D}$ charge whose origin is $\mathcal{O}_L$ symmetry.
Also, if we trivialize the $Z^{0D}$ charge of the NL (NS) below the $E_F$ in a four bands model, it opens up a gap along the NL (NS) and band structure effectively becomes a two bands system. Nonetheless, in two bands case, NLs (NSs) are not robust as shown in main paper. Now we ask a question how can we trivialize the $Z^{0D}$ charge of the NL (NS) below the $E_F$ to open up a gap. To do that the $\mathcal{O}_L$ eigen values have to be distributed as (-1,+1,-1 and +1) for the four bands above and below the $E_F$ in the a order from low energy to high energy (\ref{Zc}(a)).
For this configuration of $\mathcal{O}_L$ eigen values, the $Z^{0D}$ charge of the NLs (NSs) at $E_F$ are opposite i.e, +1 and -1 (\ref{Zc} (b)). As such the NLs are not robust against pair annihilation.
  However, such distribution of $\mathcal{O}_L$ eigen values are forbidden in KABLG under the relevant symmetries of the system. 
  In KBLG, $\mathcal{O}_L$ eigen values are distributed over the four bands as follows;
  two occupied bands take (-1,-1) and two unoccupied bands take (+1,+1) $\mathcal{O}_L$ eigen values respectively as shown in Fig.~\ref{Zc}(c). This distribution of $\mathcal{O}_L$ eigen values over the four bands does not allow to trivialize the 0D charge of NL (NS) below the $E_F$ which makes the linking structure impossible to remove (Fig.~\ref{Zc}(c,d)).
   
  Let us discuss the symmetry constraints on the nodal structure and distribution of $\mathcal{O}_L$ eigen values over the bands in KBLG under the relevant symmetries to understand this unusual robustness. 
  
KABLG possess $\mathcal{C}_3$, $\mathcal{PT}$, and two chiral symmetries $\mathcal{C}$ and $\mathcal{C}^\prime$. 
$\mathcal{C}_3$ rotation has three eigen values.
 For spinless case, they are $e^{i\frac{2\pi}{3}p}$ (p=0,1,2). 
 We denote the complex eigen values ($e^{i\frac{2\pi}{3}p}$ (p=1,2)) as $\lambda$ and $\lambda^*$. 
 Bands with complex (real) $\mathcal{C}_3$ eigen values will be called as complex (real) bands. 
 We have $[H,\mathcal{C}_3 ]=0$ and [$T,\mathcal{C}_3$]=0.
  Lets choose an eigen state $\ket{\psi}$ for which $\mathcal{C}_3 \ket{\psi} = \lambda \ket{\psi}$. 
  Then,  $\mathcal{C}_3 T\ket{\psi} = \lambda^* T\ket{\psi}$. This means $\ket{\psi}$ and its time reversed state $T\ket{\psi}$ forms a degenerate eigen space on the $\mathcal{C}_3$ rotation axises with complex rotation eigen values  $\lambda$ and $\lambda^*$. 
  However, bands with real $\mathcal{C}_3$ eigen values stay non-degenerate. 
  As a consequence, at $\Gamma$ points bands are either two-fold or singly degenerate. 
  At K point, however, bands could be  two-fold degenerate because of combine $\mathcal{PT}$ symmetry and $\mathcal{C}_3$ rotation symmetry. Please note that the two-fold degenerate points possess quantized Berry phase at K point because of the combine $\mathcal{PT}$ symmetry.
   All the bands are singly degenerate at M points as $\mathcal{C}_2$ rotation at M point allows only real eigen valued bands representation.

Also, $\mathcal{O}^{-1}_L H(k) \mathcal{O}_L = H(k)$ and $\mathcal{O}^2_L =1$. 
Also $[\mathcal{O}_L, T]=0$. 
Let, $\mathcal{O}_L \ket{\psi} = \pm \ket{\psi}$. 
Then, $\mathcal{O}_L T\ket{\psi} = \pm T\ket{\psi}$. 
This means a state and its time reversed state will have same $\mathcal{O}_L$ eigen values. 
So, on the $\mathcal{C}_3$ rotation axis, two degenerate bands which have $\lambda$/$\lambda^*$ as their $\mathcal{C}_3$ representation, possess either +,+ or -,- as their $\mathcal{O}_L$ eigen values representation.

Finally, the chiral symmetry ($\mathcal{C}$) impose the condition that a state with + (-) $\mathcal{O}_L$ eigen values at $-E$ enforce its chiral symmetric state at $E$ with - (+) $\mathcal{O}_L$ eigen values. The prove is as follows. $\{\mathcal{O}_L, \mathcal{C} \}=0$ , $\{H, \mathcal{C} \}=0$ and $[H, \mathcal{O}_L]=0$. Lets, $H \ket{\psi} = E \ket{\psi}$ then $H C \ket{\psi} = -E C\ket{\psi}$ also, $\mathcal{O}_L \ket{\psi} = \pm \ket{\psi}$ and $\mathcal{O}_L C\ket{\psi} = \mp C\ket{\psi}$. So, the chiral symmetry ($\mathcal{C}$) impose the condition that a band with + (-) $\mathcal{O}_L$ eigen value at $-E$ enforce its chiral symmetric band at $E$ to have - (+) $\mathcal{O}_L$ eigen value. Hence, for the four bands case near E$_F$, the crossing points between the Chiral partners are protected by the different quantized eigen values of $\mathcal{O}_L$.

\subsection*{2. 1D $\mathbb{Z}_2$ charge linking structure in BLG and KABLG}
In this section, we focus on the linking structure of nodal systems belonging to BDI class of $\mathcal{AZ+I}$ classification. In BDI class, a pair of NLs (NSs) at $E_F$ are linked with a NP (NL) below the $E_F$.
Let us consider $S^1$ be is a charge calculating circle which enclose the NL (NS) at the $E_F$ and $c_{1d}(S^1)$ be the corresponding 1D homotopy charge on this loop. 
$\bar{c}_{1d}(\bar{S^1})$ is the charge of NP (NL) below the $E_F$ on the loop $\bar{S^1}$ the nodal line which enclose the NP (NL) in 2D (3D).
The linking structure is present if $c_{1d}(S^1) = \bar{c}_{1d}(\bar{S^1})$ and the 0D charge of NL (NS) at $E_F$ $c_{0d}(S^0)\neq 0$. Indeed $c_{0d}(S^0)\neq 0$ because $A_F(k) \in O(N)$ under the BDI symmetries.

Let us define the off-diagonal block of the 2N bands flattened Hamiltonian inside and out side the nodal line at $E_F$  as,
\begin{equation}
A^{FB\pm}({\bf k}) = \pm \ket{u^{\uparrow}_{1\bf{k}}} \bra{u^{\downarrow}_{1\bf{k}}} + \mathlarger{\sum}_{n=2}^{N} \ket{u^{\uparrow}_{n\bf{k}}} \bra{u^{\downarrow}_{n\bf{k}}}
\end{equation}
where $A^{FB+}(\bf{k})$ ($A^{FB-}(\bf{k})$) represents the off diagonal block of the flattened Hamiltonian inside and outside the nodal loop at $E_F$. Actually $A^{FB+}({\bf k})$ and $A^{FB-}({\bf k})$ are related by the band inversion between $\ket{u_n^{occ}}$ and $\ket{u_{n+1}^{unocc}}$ states. We also define a function $A^{FB\prime}({\bf k})$ inside the nodal line on $S^{\prime 1}$ which has similar form as $A^{FB-}({\bf k})$.

$c_{1d}({S^1})$ ($\pi_1(S^1)$) is defined as the homotopy equivalence of the $SO(N)$ group.
 Now by assumption, $A^{FB-}({\bf k})$ can be continuously connected to  $A^{FB\prime}({\bf k})$. 
 So $\pi_1(S^1)$ can alternatively defined by the homotopy equivalence \cite{L2}; 
 \begin{equation}
 \pi_1(S^1) = [A_F^\prime({\bf k}):S^{\prime 1}\rightarrow SO(N)]
 \end{equation}
 where by assumption, $S^1$ and $S^{\prime 1}$  are continuously connected by smooth deformation. 
 Now we check whether the $S^{\prime 1}$ can smoothly deformable to a point or not. It is impossible if there is a nodal point below the $E_F$ where $\ket{u_n^{occ}}$ and $\ket{u_{n+1}^{occ}}$ become degenerate and $A^{FB\prime}({\bf k})$ is not well define on the deformed $S^{\prime 1}$. 
 This means the loop $S^{\prime 1}$ has an topological obstaction for the nodal point below the $E_F$.
  This argument establish the underlying linking structure between the NL (NS) at $E_F$ and NP (NL) below $E_F$. 

Let us show mathematically for a 4 bands Hamiltonian that the charge of NL (NS) at $E_F$ ($c_{1d}({S^1})$) and the charge of the NP (NL) below $E_F$ ($\bar{c}_{1d}(\bar{S^1})$) are equivalent. 
The generic form of $A^\pm({\bf k})$ for 4 band BDI Hamiltonian can be parameterized by $\theta$ and $\phi$ as \cite{L2}

\begin{equation}\label{AF}
\begin{aligned}
A^\pm({\bf k})= &
\frac{E_2 \mp E_1}{2} \begin{pmatrix}
sin  \theta({\bf k}) & cos  \theta ({\bf k})\\
cos  \theta({\bf k}) & -sin \theta ({\bf k})\\
\end{pmatrix}                             \\
& +
\frac{E_2 \pm E_1}{2}\begin{pmatrix}
cos  \phi({\bf k}) & -sin  \phi({\bf k}) \\
sin  \phi({\bf k}) & cos  \phi({\bf k}) \\
\end{pmatrix},
\end{aligned}
\end{equation}
where $\pm E_1$ and $\pm E_2$ ($0\leq E_1\leq E_2$) are the band energies.
 $A^+({\bf k})$ and $A^-({\bf k})$ are related by band inversion between highest and lowest occupied and unoccupied bands which are defined outside and inside of the nodal line. The sign of the determinant define the $\pi_0$ charge as discussed previously. 
 The flattened Hamiltonian outside the NL (NS) depends on $\theta$ only.
 As such the 1D charge on a loop outside the NL (NS) is given by 
\begin{equation}\label{1d}
c_{1d}({S^1})=\frac{1}{2\pi}\oint_{S^1} dk \cdot\nabla \theta
\end{equation}

Now we can compute the charge of the NP (NL) below the $E_F$ on a circle $\bar{S}^1$ inside the NL (NS) at $E_F$. We write an effective Hamiltonian as 

\begin{equation} \label{heff}
H_{eff}({\bf k}) = \ket{u_{1\bf k}^{occ}} \bra{u_{1 \bf k}^{occ}} - \ket{u_{2 \bf k}^{occ}} \bra{u_{2 \bf k}^{occ}}.
\end{equation}
We then find the explicate from of $\ket{u_{1 \bf k}^{occ}}$ and $\ket{u_{2 \bf k}^{occ}}$ from Eq.~\ref{AF} and use them in Eq.~\ref{heff}.
 The effective Hamiltonian possesses chiral symmetry. Rewriting the effective Hamiltonian in that chiral symmetry basis, we obtain the expression for off-diagonal block as  

\begin{equation}
A_{eff}({\bf k})=
\frac{i e^{i \theta (\bf k)}}{2} \begin{pmatrix}
-e^{i \phi (\bf k)} & 1 \\
1 & -e^{-i \phi (\bf k)}\\
\end{pmatrix}
\end{equation}

Using this $A_{eff}({\bf k})$, we can compute the charge of the nodal point as

\begin{equation}\label{winding}
\begin{split}
\bar{c}_{1d}({\bar{S}^1}) & =\frac{i}{2\pi}\oint_{\bar{S}^1} d{\bf k} \cdot Tr[A_{eff}^\dagger({\bf k}) \nabla A_{eff}({\bf k})] \\
 & = \frac{1}{2\pi}\oint_{S^1} d{\bf k} \cdot{\bf \nabla} \theta
\end{split}
\end{equation}

Since $\theta({\bf k})$ is continuous across the NL (NS) at $E_F$, Eq.~\ref{winding} and Eq.~\ref{1d} represent the equivalence between the charge $c_{1d}({\bar{S}^1})$ and $\bar{c}_{1d}({\bar{S}^1})$.

\subsubsection*{a. Four band case and application to BLG}
\begin{figure*}[t!]
\centering
\includegraphics[width=1.0\textwidth]{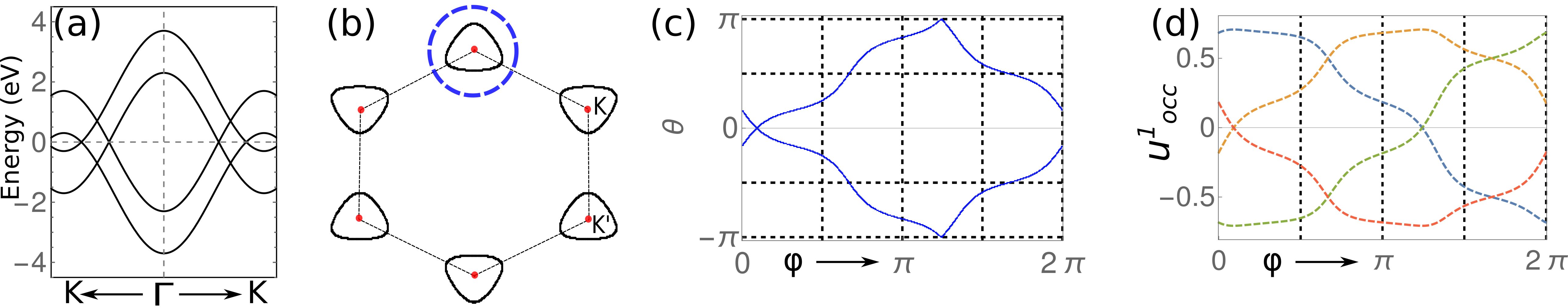}
\caption{(a) Band structure of AA-stacked bilayer graphene for $t=1$, $t_c=0.7$. (b) Corresponding nodal structure. Black loops around the K/K$^\prime$ points represent the nodal line at $E_F$ which are linked with nodal point (represented by red dots) below the $E_F$. The blue dashed line is the charge calculating loop. (c) The phases of the eigen values of the  $A_{FB}(k) \in SO(2)$ are plotted on the blue loop to compute the $\pi_1$ charge of the nodal line at $E_F$. (d) Orientation of wavefunction $\ket{u^1_{occ}}$ on the close circle surrounding the nodal point below the $E_F$.}
\label{BLG}
\end{figure*}

A four bands Hamiltonian for AA-stacked  bilayer graphene (ABLG) can be written as,
\begin{equation}
 H_{ABLG}({\bf k})=
 \begin{pmatrix}
H_{SLG}({\bf k}) & t_c                 \\
 t_c & H_{SLG}({\bf k})                \\
\end{pmatrix}
\end{equation}
where $H_{SLG}({\bf k})$ represents the Hamiltonian for single layer graphene which is given by  
$\begin{pmatrix}
0 & f({\bf k})            \\
f^{\dagger}({\bf k}) & 0 \\
\end{pmatrix}$
where $f({\bf k}) = -t(e^{-ik_x} +2 e^{i \frac{k_x}{2}}cos(\frac{\sqrt{3} k_y}{2}))$. 
$H_{SLG}$ represents the single layer graphene Hamiltonian and $t_c$ is the coupling parameter between the two layers. 
The band and nodal structures are shown in Fig.~\ref{BLG}(a,b).
 Nodal NLs and NPs are represented by black and red color around and at the K/K$^\prime$ points. 
 Blue color dashed loop is the charge calculating manifold around the nodes. 
 The Hamiltonian respect $\mathcal{T}$ and $\mathcal{I}$  and chiral symmetry. As such it falls into the class BDI. 
 After we choose a representation of the symmetries as $\mathcal{TI}=\mathcal{K}$ and $\mathcal{C}=\sigma_z$, where $\mathcal{K}$ is the complex conjugation operator, the Hamiltonian becomes the off-diagonal block form
  $H_{ABLG}({\bf k})=\begin{pmatrix}
0&&A({\bf k})  \\   A^T({\bf k}) &&  0
\end{pmatrix}$. 

The sign of $\mathrm{det}A({\bf k})$ can be changed by band inversion \cite{L2}.

We have discussed in previous section that the 0D $\pi_0$ charge of the nodal line is defined by the properties of the two disconnected components of the $O(N)$ group. i.e, the sign of $det A({\bf k})$ changes inside and outside of the nodal line at $E_F$. 
We verify this for the BLG Hamiltonian which confirms the $\pi_0$ charge of the nodal line at $E_F$.

However, to calculate the 1D $\pi_1$ charge of the nodal loop at the $E_F$, we parameterize the flattened Hamiltonian by angle $\phi$ and then plot the phase of eigenvalues of the off-diagonal block $A({\bf k})$.
 We also use a  mirror symmetric transformation on $A({\bf k})$, if $det A({\bf k})<0$, to make it an element of $SO(N)$ group. Being an element of $SO(2)$ (in case of BLG), $A({\bf k})$ has two eigenvalues in the form $exp(\pm i\phi)$.
  Therefor the phase of the eigen values of $A({\bf k})$ (represents the Euler angle ($\theta$) of $SO(2)$) represents the topological charge of $S^1$ which have been shown in Fig.~\ref{BLG}(c). 
  The phase change takes place for $\phi$ changes from 0 to $2\pi$, indicating the nontrivial charge of nodal line at $E_F$.

The 1D charge of the nodal point below the $E_F$ can be checked by investigating the orientation of the wave functions.
 Below the $E_F$, the nodal point is protected by $\mathcal{PT}$ symmetry which enforce nontrivial Berry phase corresponding to the node. 
 To check the 1D charge, we plot the components of $\ket{u^1_{occ}}$ of the Hamiltonian, parameterized by $\phi \in [0,2\pi]$ as shown in Fig.~\ref{BLG}(d).
 The components acquire a phase which indicates the nontrivial charge of nodal point below the $E_F$.
  
 Such doubly charge ($\pi_0$ and $\pi_1$) nodal line at $E_F$ and nontrivial nodal point below the $E_F$ arise together and form a robust linking structure. 
The linking structure in BLG is that the $\pi_1$ charge of a nodal line at $E_F$ is non-trivial if and only if there is a nodal point, which is below the $E_F$, inside the nodal line \cite{L2}.

\subsubsection*{b. 2N (N$>$2) band case and application to KABLG}
In this section, we compute the topological charges and check the robustness of the nodal structure of KABLG under strong deformations. 
Let us first formally define the $\pi_1$ charge of BDI class for 2N (N$>$2) bands system. For a 2N system, the $\pi_1$ charge is defined as the homotopy equivalence class of the $A^{FB}({\bf k}) \in  SO(N)$ where $A^{FB}({\bf k})$ is N$\times$N  off-diagonal block of the flattened Hamiltonian. 
The element of $SO(N)$ group can be generated by using N(N-1)/2 number of $SO(N)$ generators which are defined as 

\[
    [L_{ij}]_{nm}= 
\begin{cases}
  -1,   & \text{if } i=n, j=m \\
   1,   & \text{if } i=m, j=n \\
    0,              & \text{otherwise}
\end{cases}
\]

 where $i<j$. $A^{FB}$ can be written as\cite{L2}

\begin{equation} \label{genexp}
A^{FB} = M_N.M_{N-1}.M_{N-2} ....M_2 
\end{equation} 
where matrix $M_N$ is given as

\begin{equation} \label{eq1}
\begin{split}
M_N & = e^{\alpha_1 L_{12}}.e^{\alpha_2 L_{23}} .....   e^{\alpha_{N-1} L_{N-1,N}} \\
M_{N-1} & = e^{\beta_1 L_{12}}.e^{\beta_2 L_{23}} .....   e^{\beta_{N-2} L_{N-2,N-1}} \\
M_{N-2} & = e^{\gamma L_{12}}.e^{\gamma_2 L_{23}} .....   e^{\gamma_{N-3} L_{N-3,N-2}} \\
.\\
.\\
.\\
M_{2} & = e^{\delta_1 L_{12}}\\
\end{split}
\end{equation}

Here the matrix exponential $e^{\alpha L_{i,j}}$ of a generator $L_{i,j}$ represents the element of Lie group $SO(N)$, i.e, $M_i \in SO(N)$ ($i=N, N-1,....,2$).

The $\pi_1$ charge of the nodal line is defined as 
\begin{equation}
\pi_1(S^1) = [A^{FB}({\bf k}):S^1\rightarrow SO(N)] 
\end{equation} 

To determine the $\pi_1$ charge, we lift the close loops from $SO(N)$ to its \textit{universal covering} space i.e, $Spin(N)$. $Spin(N)$ is the doubly covering space of $SO(N)$ which is also \textit{simply connected}. If the image of lifted loop in $Spin(N)$ is open (close), the element of $\pi_1$ in $SO(N)$ correspond to non-trivial (trivial) topology. 
The lifting can be achieved by changing the $SO(N)$ generators in Eq.~\ref{genexp} into $Spin(N)$ generators\cite{bzduvsek2019nonabelian}.

Let us now apply the above idea on the 12-bands KABLG graphene Hamiltonian. As we discussed in the main paper the KABLG can be in a phase with $t_2<t_1$ (anti-Kekul\'{e} distortion) or $t_2>t_1$ (Kekul\'{e} distortion). In both cases, the nonzero interlayer hopping ($t_3$) pushes the gapped band structure of each textured graphene layer, upward and downward in energy, respectively, which generates linked NLs.
Here, we discuss the robustness in the both case explicitly under strong deformation.

\begin{figure}[t!]
\centering
\includegraphics[width=0.5\textwidth]{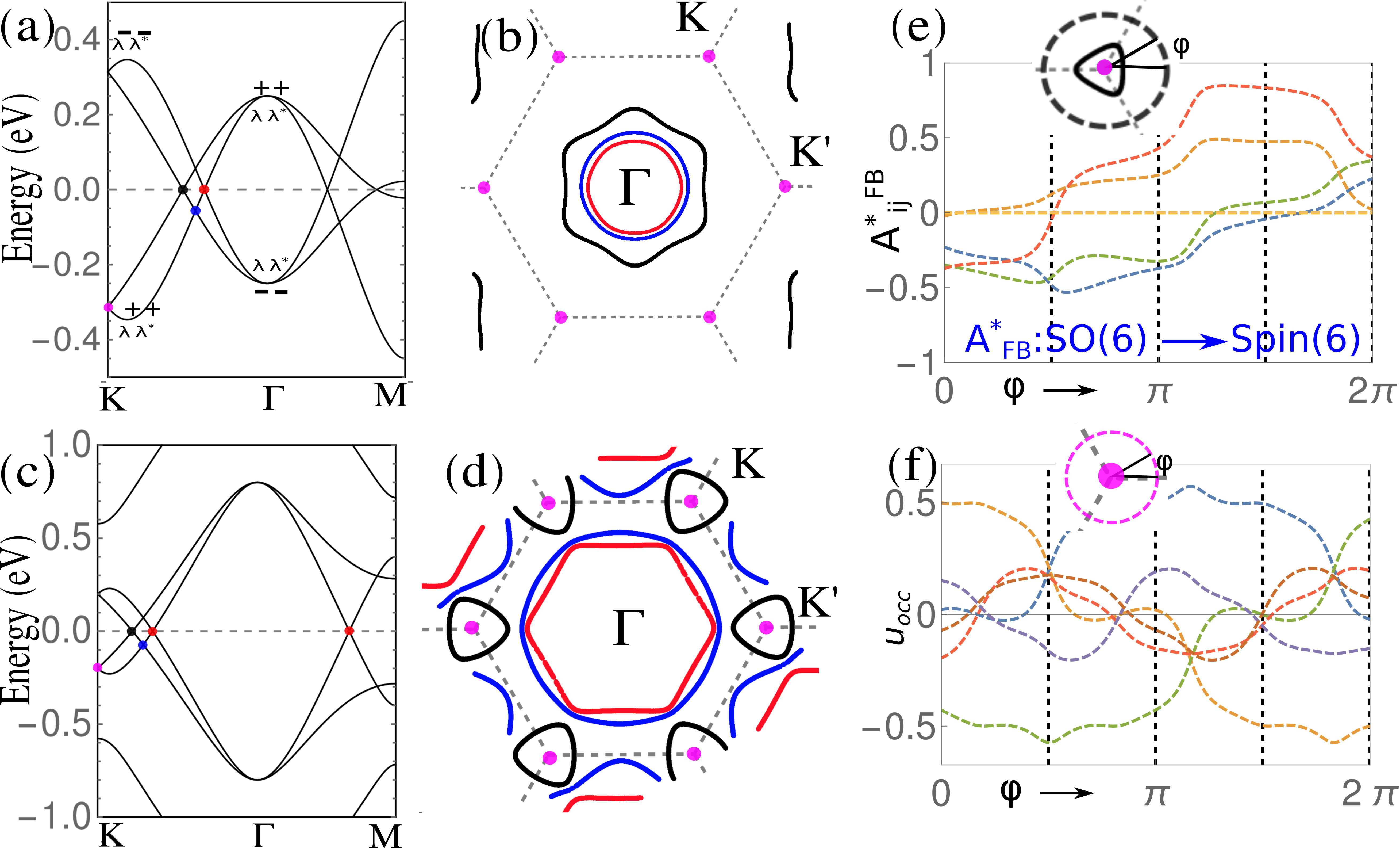}
\caption{(a) Band structure of KBLG using $t_1=1$, $t_2=0.6$ and $t_3=0.8$. 
         (b) Corresponding nodal structure. $\lambda$,$\lambda^*$ and +,- represents the $\mathcal{C}_3$ and $\mathcal{O}_L$ eigen values. 
        (c,d) Band and nodal structures after Lifshitz transition (LT) for stronger inter layer coupling $t_3=1.2$. 
        (e,f) $Z_2$ type $\pi_1$ charges for NLs at $E_{F}$ and NPs below $E_{F}$ after LT around and at $E_{F}$. 
        The matrix elements of A$^{*FB}$ and the components of $u^{occ}$ states are represented by different colors on the charge calculating loops in (e) and (f).
Red and black NLs lie at the $E_F$, blue NL lies below the $E_F$ and magenta NP lies below the $E_F$ in (b) and (d).}
\label{4bands}
\end{figure}

As shown in Fig.~\ref{4bands} (a,b), the band and nodal structures of the KABLG for anti-Kekul\'{e} distortion with parameters $t_1=1$, $t_2=0.6$ and $t_3=0.8$. Two concentric NLs at $E_F$ are doubly charged and linked with another NL below $E_F$.
As the two NLs at $E_F$ carry the same $Z^{0D}$ charge, they cannot be pair-annihilated through merging as discussed before.
As $t_3$ increases, the size of the outer NL at $E_F$ becomes larger, and after touching the BZ boundary, it splits into two NLs encircling ${\bf K}$ and ${\bf K^{\prime}}$ (Fig.~\ref{4bands} (c, d)).
Interestingly, each of this NL develops another type of the linked nodal structure with a NP below $E_F$.
As each NL at ${\bf K}$ or ${\bf K^{\prime}}$ is doubly charged, it cannot be annihilated separately as discussed below.

  Fig.~\ref{4bands} (c,d) show the band and nodal structures for $t_1=1$, $t_2=0.6$ and $t_3=1.2$. For these parameters, two NLs at ${\bf K}$ or ${\bf K^{\prime}}$ at $E_F$ (shown in black color) are linked with NPs below the $E_F$. The charges of the nodes are discussed as follows. 
To compute the 0D and 1D $Z_2$ homotopy charges, we adopt the spectral flattening technique. Because of the $\mathcal{PT}$ and $\mathcal{C}$, the space of flattened KABLG Hamiltonian belongs to $O(6)$ as the off-diagonal block $A^{FB}({\bf k})$ of the flattened Hamiltonian is 6$\times$6 real matrix. The NLs at $E_F$ have both 0D and 1D $Z_2$ charges. The $\pi_0$ charge of the NLs can be represented by the two disconnected component of $O(6)$ group, e.g, the space with determinant $\pm 1$ of the $O(6)$ as discussed previously. In fact, $\pi_0$ charge represents the band inversion. The $\pi_1$ charge is defined on the 1D manifold (black dashed loop that encircle the NL [see Fig.~\ref{4bands}(e)]) which is given by the relation, $\pi_1(S^1) = [A^{FB}({\bf k}):S^1\rightarrow SO(6)]$. $SO(6)$ group possess 15  parameters which are nothing but the generalized Euler angles\cite{L2}. To determine the $\pi_1$ charge, we lift the close loops from $SO(6)$ space to its universal covering space $Spin(6)$. $Spin(6)$ is the doubly covered space of $SO(6)$, which is also simply connected. If the image of lifted loop (i.e, \textit{preimage}) in $Spin(6)$ space is open (close), the element of $\pi_1$ in $SO(6)$ correspond to non-trivial (trivial). Because a close loop can be shank to a point as the covering space is simply connected. However, it is impossible for a open path \cite{bzduvsek2019nonabelian}. Fig.\ref{4bands}(e) represents the lifted open path (because the initial and final values are different) on the covering space which suggest the non-trivial $\pi_1$ charge of the NL at $E_{F}$. The NL at $E_{F}$ is linked with the nodal point below the $E_{F}$ which also possess non-trivial Berry phase which can be equivalently represented by computing the orientation of the real wavefunctions as shown in Fig.~\ref{4bands}(f)

\begin{figure}[t!]
\centering
\includegraphics[width=0.50\textwidth]{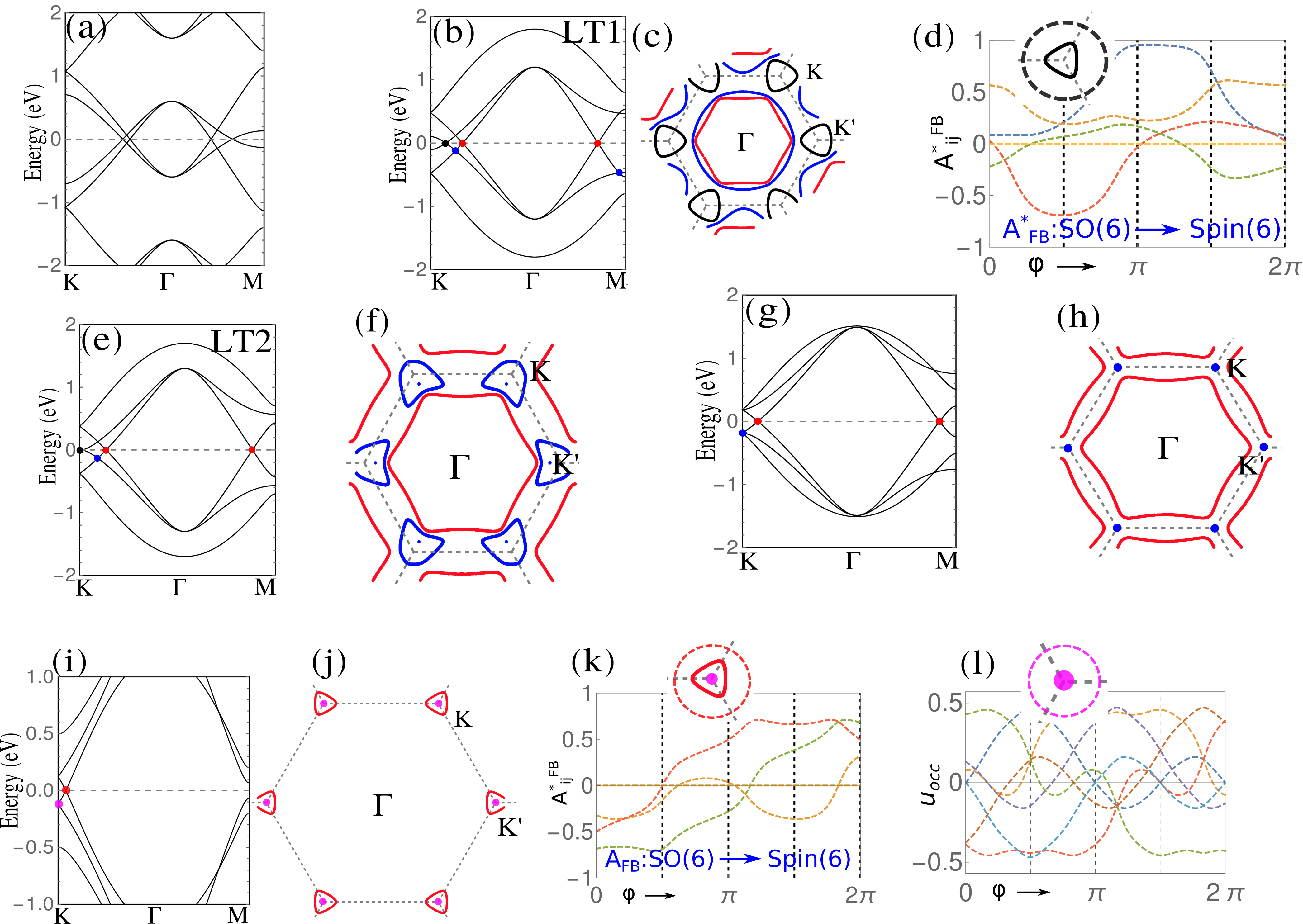}
\caption{Linking structure for Kekul\'{e} modulation i,e for $t_1<t_2$.  
    (a) Band structure for $t_1=1$, $t_2=1.5$ and $t_3=1.1$. NL-NL linking structure is enforced by 0D $Z^{0D}$ charge around $\Gamma$ point.
    (b,c) Band and nodal structure after the LT for larger interlayer coupling at $t_3$=1.7. 
   (d)Trivial $\pi_1$ charges represents non linking structure.
   (e,f) Band and nodal structure for further increased $t_3 =2.0$. The black NLs at $E_F$ shrink to points and blue NL below $E_F$ shifts around the K/K$^\prime$ points through a LT. 
   (g,h) Black NLs are annihilate and blue NLs shrink to points for increased $t_3$.   
   (i,j) Band and nodal structure for $t_3=2.3$ with NLs at $E_F$ which are linked with NPs below $E_F$.  
   (k,l) $Z_2$ 1D non-trivial charges for NL and NP confirms the linking structure.}
\label{6bands}
\end{figure}

In case of Kekul\'{e} modulation (i.e, $t_2>t_1$), high energy bands hybridize with four bands near $E_{F}$. 
Such hybridization effects the robustness of the linking structure. Figure.~\ref{6bands}(a) shows the bands structure for $t_1=1$, $t_2=1.5$ and $t_3=1.1$. For these parameters, the nodal structure around the $\Gamma$ is exactly similar to the anti-Kekul\'{e} modulation (i.e, $t_2<t_1$).
 However, increasing inter layer coupling of $t_3=1.7$, the nodal structure goes through a LT as shown in Fig.~\ref{6bands}(b,c). 
 Please note that the linking structure is broken. 
 Because unlike the anti-Kekul\'{e} modulation case, 
 now there are no NPs at ${\bf K}$ and ${\bf K^{\prime}}$ below the $E_{F}$ to establish a $Z_2$ charged linking.
  The absence of $Z_2$ linking can be confirmed by checking the lifted $\textit{preimage}$ in the $Spin(6)$ covering space as shown in Fig.~\ref{6bands}(d). 
  Interestingly, if we further increase the $t_3$, the nodal structure evolves through further LT as shown in Fig.~\ref{6bands}(e,f,g,h). Finally  
 at $t_3=2.3$ a new NL-NP linking arises around ${\bf K}$ and ${\bf K^{\prime}}$ points as shown in Fig.~\ref{6bands}(i,j).
   Corresponding non-trivial charges of the linking nodes further supports this arguments (see Fig.~\ref{6bands}(k,l)).
    Now, the linked nodes in Fig.~\ref{6bands}(i,j) can only be annihilated pairwise.

\section*{S5. DFT computational details}

First principle calculations were carried out using projector augmented wave (PAW)\cite{PEBLOCH1994} formalism based on Density Functional Theory (DFT) as implemented in the Vienna Ab Initio Simulation Package (VASP).\cite{Hafner1993,Joubert1999} The generalized-gradient approximation by Perdew-Burke- Ernzerhof (PBE)\cite{Joubert1999} was employed to describe the exchange and correlation. Force (energy) criterion was set upto 0.01 eV/\AA(10$^{-6}$ eV). An energy cut off of 500 eV is used to truncate the plane-wave basis sets. The Brillouin zone (BZ) is integrated using 8$\times$8$\times$4 $\Gamma-$centered k-mesh. DFT-D2 method has been applied for Van der Waals forces correction. Tight-binding (TB) Hamiltonians were constructed using maximally localized Wannier functions (MLWFs)\cite{mlwf1,mlwf2,mlwf3} obtained from wannier90 package.\cite{w90} The topological properties including  surface states (SSs) were analyzed based on the iterative Green's function\cite{greenfn1,greenfn2,greenfn3} method implemented in Wannier-Tools package.\cite{WTools}

\begin{figure}[t!]
\centering
\includegraphics[width=0.5\textwidth]{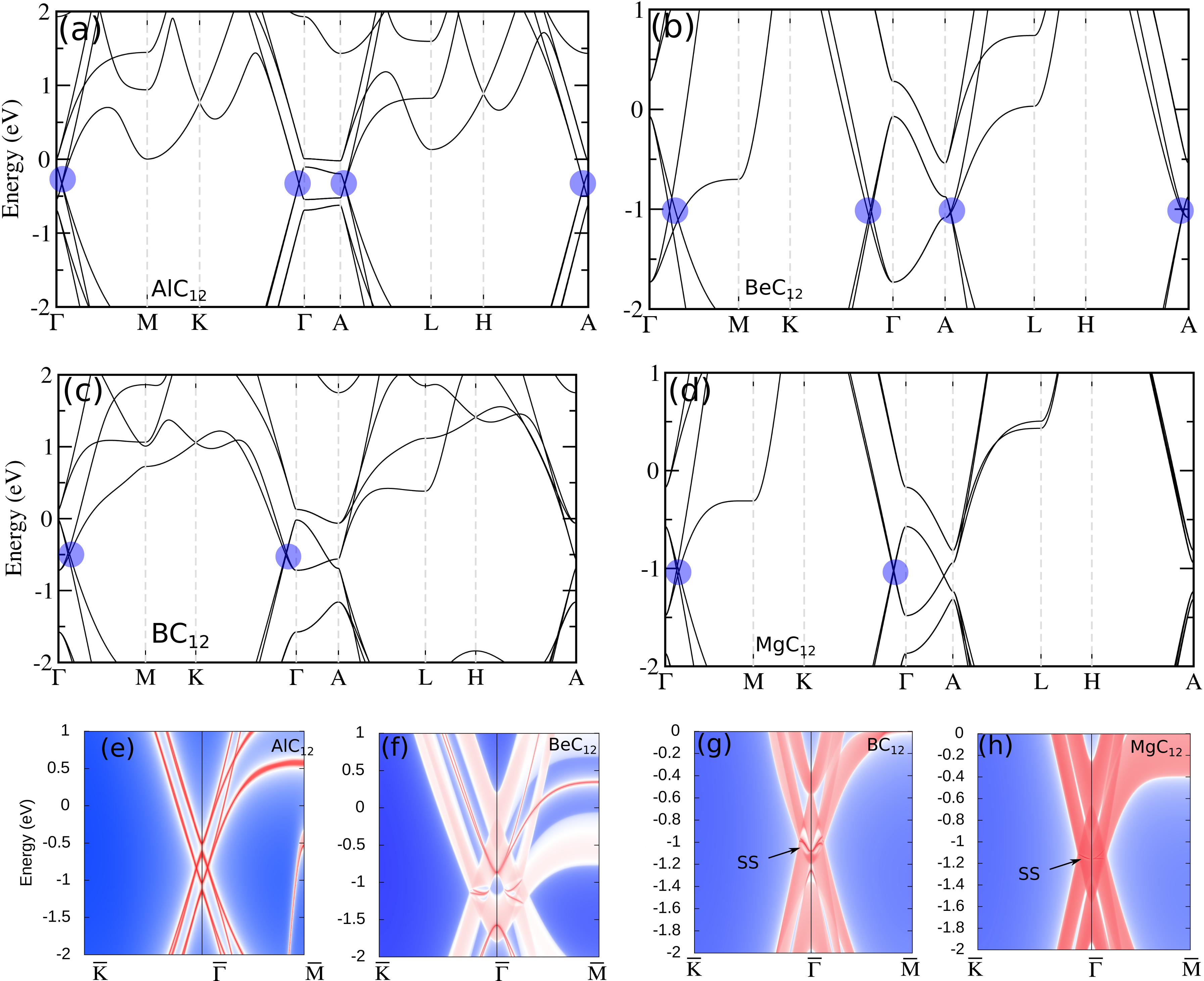}
\caption{Double band inversion occurs at every points along $\Gamma$-A direction in (a) AlC$_{12}$, (b) BeC$_{12}$. 
 In (c) BC$_{12}$, (d) MgC$_{12}$, double band inversion occurs around $\Gamma$ point and show nodal line degeneracies. 
 The blue shaded region indicated the nodal degeneracies.
  Surface states of (e) AlC$_{12}$, (f) BeC$_{12}$, (g) BC$_{12}$ and (h) MgC$_{12}$ on (001) surface. 
  Non-trivial Berry phase driven nodal line flat band surface states are indicated by arrows in (g,h).}
\label{material}
\end{figure}

\subsection*{1. Additional materials and surface state calculation}

We compute band structures of similar compounds such as AlC$_{12}$, BeC$_{12}$, BC$_{12}$ and MgC$_{12}$. These materials have similar nodal structures like ZnC$_{12}$ or LiC$_{12}$. For instance, AlC$_{12}$, BeC$_{12}$ show double band inversion at every points along the k$_z$ direction ($\Gamma$-A path) as shown in Fig.~\ref{material}(a-b). These are the possible candidate for $\textit{linked}$ nodal surfaces. However, in BC$_{12}$ and MgC$_{12}$ double band inversion occurs around the $\Gamma$ point (as shown in Fig.~\ref{material}(c,d)) which forms the nodal line structure similar like LiC$_{12}$.
 We have also computed surface states for all the predicted compounds as shown in Fig.~\ref{material}(e-h). As we have discussed that AlC$_{12}$, BeC$_{12}$ possess band inversion at every point along k$_z$ axis and these compounds might show cylinder like nodal surface. These nodal cylinder can be consider as composition of even number of nodal lines (one at k$_z$=0, another at k$_z = \pi$ and all other nodal line at any k$_z$ will have a time reversal partner at -k$_z$).  As such net Berry phase correspond to these nodal cylinder is trivial. Hence they do not show flat band surface states. However, BC$_{12}$ and MgC$_{12}$ possess flat band surface states which comes from the non-trivial Berry phase of the nodal lines at k$_z$=0 plane. Here, we mention that none of these system strictly maintain the chiral symmetry of the systems. In fact, no real compound respect such symmetry perfectly. Note that surface states in BC$_{12}$ and MgC$_{12}$ are not completely flat bands. This is because strong breaking of chiral symmetry in these systems.
